\documentclass[twocolumn,prb,showpacs]{revtex4}
\usepackage{psfrag}
\usepackage{graphicx}
\usepackage{amsmath}
\usepackage{amssymb}
\usepackage{eufrak}
\usepackage[ansinew]{inputenc}
\usepackage{url}

\begin{document}

\title{The Higgs mass derived from the U(3) Lie group}

\author{Ole L. Trinhammer*, Henrik G. Bohr, Mogens Stibius Jensen}

\address{Technical University of Denmark, \\
DK-2800 Kongens Lyngby, Denmark\\
*corresponding author}

\begin{abstract}
The Higgs mass value is derived from a Hamiltonian on the Lie group U(3) where we relate strong and electroweak energy scales. The baryon states of nucleon and delta resonances originate in specific Bloch wave degrees of freedom coupled to a Higgs mechanism which also gives rise to the usual gauge boson masses. The derived Higgs mass is around 125 GeV. From the same Hamiltonian we derive the relative neutron to proton mass ratio and the N and Delta mass spectra. All compare rather well with the experimental values. We predict scarce neutral flavor baryon singlets that should be visible in scattering cross sections for negative pions on protons, in photoproduction on neutrons, in neutron diffraction dissociation experiments and in invariant mass spectra of protons and negative pions in B-decays. The fundamental predictions are based on just one length scale and the fine structure constant. More particular predictions rely also on the weak mixing angle and the up-down quark flavor mixing matrix element. With differential forms on the measure-scaled wavefunction, we could generate approximate parton distribution functions for the u and d valence quarks of the proton that compare well with established experimental analysis.

\end{abstract}

\maketitle

\section*{1. Introduction}
\label{sec1}
Ever since the proposal of the so-called Higgs mechanism fifty years ago \cite{EnglertBrout,HiggsSep1964,HiggsOct1964,GuralnikHagenKibble,Higgs1966} and, especially after the experimental findings and confirmations of the Higgs particle during the last two years \cite{ATLASsep2012,CMSsep2012,CMSjul2014,ATLASjun2014}, the big question is how to calculate it's mass - because the standard model did not contain a recipe for that. To remedy this, we make a step towards a unification of the quantum chromo dynamics of strong interactions with the quantum flavor dynamics of electroweak interactions. To make the step, we digress from quantum field theory into a common $U(3)$ configuration space where color and flavor are intermingled. At first sight this might seem confusing, but we shall show how one can project out both quark and gluon fields with the usual transformation properties.

The standard model contains quite a few unexplained parameters such as the six quark mass parameters, the three angles and one phase of the Cabibbo-Kobayashi-Maskawa (CKM) mixing matrix\cite{RPP2012CKMmatrix}, the several coefficients and exponents of each parton distribution function and a similar wealth for the six leptons. The strong and electroweak interactions are described by seemingly independent gauge groups $SU(3)$ and $U(2) \cong SU(2)\times U(1)$ \cite{FlorianScheckU2}. Baryons feel both interactions wherefore we seek a description from a common Lie group background. The simplest choice is $U(3)$ which contains $SU(3), SU(2)$ and $U(1)$ as exemplar subsgroups.  With this choice we can reduce considerably the number of parameters needed to describe baryon mass spectra and the Higgs mass. We stress that the group $U(3)$ is generated from three parametric momentum operators, three parametric angular momentum operators and three remaining Runge-Lenz-like operator components to connect the algebra. The six latter can be seen as intrinsic editions of the generators of the Lorentz algebra\cite{Schwartz2014}. With three dimensions in laboratory space $\mathbb{R}^3$, the group manifold $U(3)$ therefore becomes the natural choice for intrinsic degrees of freedom that can be kinematically excited from laboratory space. The dynamics of the intrinsic degrees of freedom is projected back to laboratory space in the shape of quantum fields of various structures depending on the projection base chosen. A mixing between such projections becomes natural when one considers that the related subgroups are intermingled in the common $U(3)$ configuration. This conception may open for a derivation of CKM-matrix elements although that is far beyond the scope of the present work. We shall, however, make a first step to correlate the strong and electroweak interactions of baryons, namely in a derivation of the Higgs mass.

\begin{figure} 
\begin{center}
\includegraphics[width=0.45\textwidth]{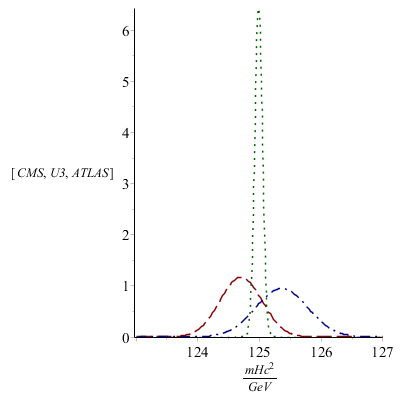}
\caption{Gaussian Higgs mass distributions as observed by the CMS collaboration (dashed, red) \cite{CMSjul2014} and the ATLAS collaboration (dashdotted, blue) \cite{ATLASjun2014} compared with the theoretical result (dotted, green) in (\ref{eq:improvednumerics}) from a common Lie group perspective for strong and electroweak interactions. The curve widths represent the standard deviations of the respective mass peak determinations and not the resonance width which is much smaller \cite{CaolaMelnikov2013}.}
\label{fig:HiggsMassGaussians}
\end{center}
\end{figure}

In this paper, we derive fundamental mass values for Higgs and gauge bosons and report on mass values for the $N$ and $\Delta$ baryon spectrum with dynamics described from a Lie group perspective. The derived Higgs mass around 125 GeV corresponds rather well to the recent experimental results \cite{CMSjul2014,ATLASjun2014} as seen in Fig. \ref{fig:HiggsMassGaussians} and is based on just one dimensionful parameter in such a way that the ratio between the Higgs mass and the electron mass, apart from mathematical constants, contains only the fine structure constant. Of other papers analyzing the Higgs mass are composite models like the one by Dhar, Mandal and Wadia \cite{DharMandalWadia}  related to the Gross-Neveu model \cite{GrossNeveu} with a Nambu-Jona-Lasinio type \cite{NambuJonaLasinio} four-fermion coupling leading to a dynamically generated Higgs mass without the need for an {\it a priori} Higgs potential.

One of us has previously introduced the Lie group $U(3)$ as configuration space \cite{TrinhammerEPL102,TrinhammerHiggsPreprint}. It contains the usual gauge groups  $SU(3)$ of strong interactions and $SU(2)\times U(1)$ of electroweak interactions. The essential frame to be adopted here complies with local gauge symmetry when the intrinsic Lie group dynamics is projected to laboratory space.  Each point $P(x,y,z)$ in space is equipped with an intrinsic $U(3)$ configuration space in which the fundamental dynamics is formulated with $u=e^{i\chi}\in U(3)$ as configuration variable. Thus our configuration space is orthogonal to the space-time manifold of the laboratory space. The closest analogue we can think of is that of intrinsic spin. In the present case the intrinsic space contains both color, spin, isospin and hypercharge degrees of freedom. We thus can capture both the strong and electroweak sections of baryon phenomena. A major  motivation is to reduce the number of ad hoc mass parameters in baryon phenomena relative to the standard model. As a benefit of intermingling the gauge groups of the standard model in a common intrinsic space, the parameters in the Higgs potential and the electroweak energy scale are determined from the relation to the intrinsic baryon potential - and the missing resonance problem in baryon spectroscopy vanishes. We do not expect to capture the meson sector since mesons are interaction quanta, i.e. field constructions in laboratory space.

\begin{figure} 
\begin{center}
\includegraphics[width=0.45\textwidth]{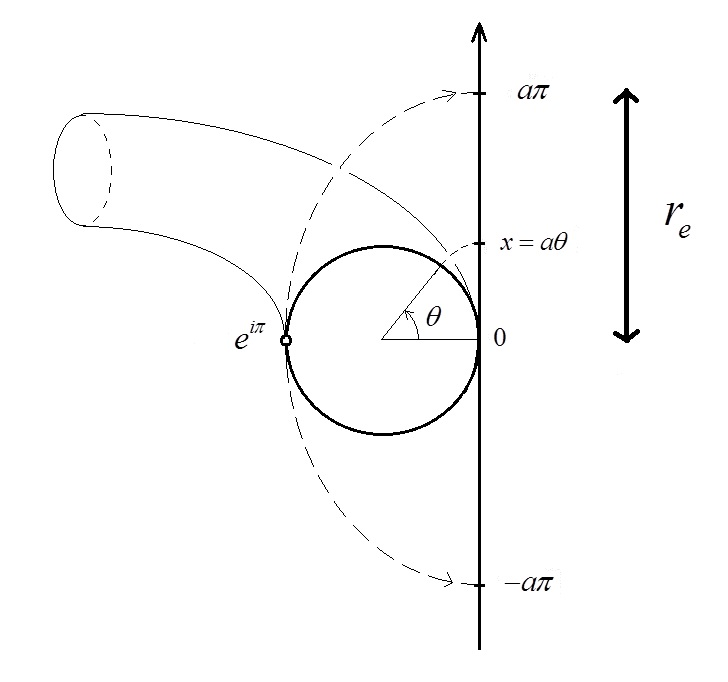}
\caption{Projection of the Lie group configuration space to the algebraic parameter space  \cite{TrinhammerEPL102}. The algebra approximates the group in the neighbourhood of the origo. The projection is scaled by the classical electron radius $r_{\rm e}$ as a measure for the extension of the charge "scar" created in the neutron decay. This corresponds excellently to the measured value for the neutron to electron mass ratio, see Fig. \ref{fig:memnConvergence}.}
\label{fig:ProjectionGroupAlgabra}
\end{center}
\end{figure}

\begin{figure}
\begin{center}
\includegraphics[width=0.45\textwidth]{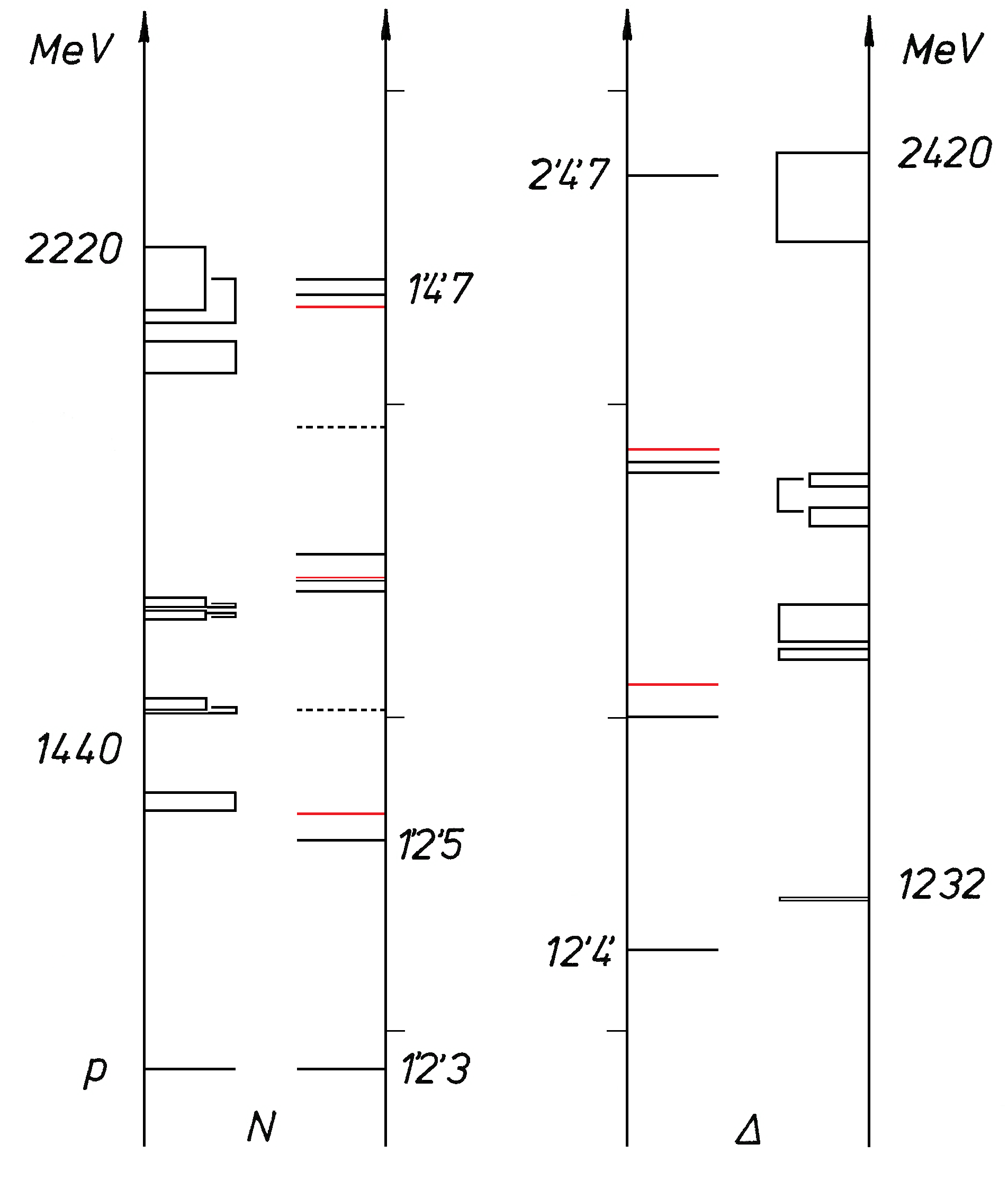}
\caption{All observed four star $N$ and $\Delta$ baryons (boxes) compared with approximate predictions (black, red and dashed lines) from eq. (\ref{eq:SeperableSchroedinger}). The dashed lines represent neutral flavor singlets, particular for the present model. The red lines mark states with augmented contribution in level 3. The boxes indicate the experimental range of pole positions \cite{RPP2012}, not the resonance widths which are much larger. We have made no estimate of mass shifts due to strong coupling to decay channels \cite{Hoehler}. Digits at selected predictions are parametric labels $p, q, r$ based on Table \ref{tab:paraeig}. Note the fine agreement in the grouping and the number of resonances in both sectors with no missing resonance problem as opposed to ordinary quark models, see Fig. \ref{fig:missingResonances}.}
\label{fig:NDeltaSpectrum}
\end{center}
\end{figure}

Our basic frame is a Hamiltonian structure on the Lie group $U(3)$ as a configuration space for baryons. We consider baryons as stationary states with masses $mc^2=E$ determined as eigenvalues of \cite{TrinhammerEPL102,KogutSusskind,Manton,Trinhammer1983}
\begin{equation} \label{eq:schroedinger}
   \frac{\hbar c}{a}\left[-\frac{1}{2}\Delta+\frac{1}{2}\rm{Tr}\ \chi^2\right]\Psi(u)=E\Psi(u)
\end{equation}
where $\Lambda\equiv\ \hbar c/a\approx 214.27\ \rm{MeV}$ is our energy scale factor corresponding to a length scale $a$. Note that $\Lambda$ corresponds to the QCD energy scale factor\cite{RPP2012p131}, e.g. $\Lambda_{\overline{\rm MS}}^{(5)}=213\pm 8\ \rm MeV$ and is of the order of the pion decay constant\cite{WeinbergQToFIIp186} $F_\pi= 184\ \rm MeV$. The latter is common for setting the scale in different phenomenological models \cite{BohrProvidenciaProvidencia2005,DiakonovPetrovPolyakov}. For our $\Lambda$ the length scale $a$ was explicitly related\cite{TrinhammerEPL102} to the classical electron radius\cite{Heisenberg,LandauLifshitz,RPP2012p107} $r_e=e^2/(4\pi\epsilon_0 m_e c^2)=\alpha\hbar c/(m_ec^2)$  by a mapping $\pi a=r_e$, between real parameter space and toroidal angles in the Lie group, see Fig. \ref{fig:ProjectionGroupAlgabra}. 
The $\Lambda$ above is calculated from a fine structure constant taken at nucleonic energies \cite{RPP2012p137} $\alpha^{-1}(m_\tau)=\alpha^{-1}(1.77\rm\ GeV)=133.471$. In stead of the Manton-inspired\cite{Manton} potential $\frac{1}{2}\rm{Tr}\ \chi^2$ acting on the generators of the configuration variable we could have chosen a Wilson-inspired\cite{Wilson,Wadia} potential $3-\frac{1}{2}{\rm{Tr}}(u+u^\dagger)$ taking the trace directly on the configuration variables themselves. Both agree in the neighbourhood of the origo $e=I$ of $U(3)$ but differ for larger deviations of the configuration variable. Both will yield the same Higgs and gauge boson masses but differ in the baryon mass spectra because the Higgs mass is determined from the shared second order term near potential minima, where the algebra approximates the group, whereas the baryon states occupy all of the intrinsic geometry. Therefore the Manton-like potential better reproduces the baryon spectrum, see Fig. \ref{fig:NDeltaSpectrum}. We also prefer the Manton-like potential because, in the parametrization of the configuration space, it represents the Euclidean measure folded onto the group manifold \cite{HansPlesnerJacobsen}. Note finally, as the configuration space is truly intrinsic, relativity only comes into play once the inherent dynamics in (\ref{eq:schroedinger}) is projected to space as when the parton distribution functions in Fig. \ref{fig:partonDistribution} below were derived in ref. \cite{TrinhammerEPL102}.

In Sec. 2 we describe the model. In Secs. 3 and 4 we carry through particular solutions for the baryon spectrum and discuss experimental predictions for unconventional baryon singlets. In Sec. 5 we describe projections to laboratory space where quantum fields resurface. In Sec. 6 we state a relation between strong and electroweak configurations and derive a Higgs mass. In Sec. 7 we relate to standard results for the vector gauge bosons. In Sec. 8 we give remarks on interpretations and in Sec. 9 we suggest lines of future study.

\section*{2. Unfolding the model}
\label{sec2}

One may consider the basic equation (\ref{eq:schroedinger}) as an effective theory inspired by lattice gauge theory \cite{KogutSusskind, Manton, Trinhammer1983}. However, we prefer to present it as detached from this framework such that the configuration space and the space-time manifold orthogonal to it are both continuous. 
In the basic equation (\ref{eq:schroedinger})
\begin{equation}
   \frac{\hbar c}{a}\left[-\frac{1}{2}\Delta+\frac{1}{2}\rm{Tr}\ \chi^2\right]\Psi(u)=E\Psi(u), \nonumber
\end{equation}
the wavefunction $\Psi$ is a function of $u=e^{i\chi}\in U(3)$. Analogously to the euclidean Laplacian in polar coordinates
\begin{equation} 	\label{eq:radialLaplacian}
  \Delta_{e,polar}=\frac{1}{r^2}\frac{\partial}{\partial r}r^2\frac{\partial}{\partial r}-\frac{1}{r^2}{\bf L}^2,
\end{equation}
here the Laplacian $\Delta$ in (\ref{eq:schroedinger}) can be parametrized in a polar decomposition \cite{TrinhammerOlafsson} ($\hbar =1$)
\begin{equation}    \label{eq:Laplacian}
 \Delta= \sum^3_{j=1} \frac{1}{J^2} \frac{\partial}{\partial \theta_j} J^2 \frac{\partial}{\partial \theta_j} - \! \! \! \sum^3_{\substack{ i <  j \\ k\neq i,j}} \frac{K^2_k + M^2_k}{8 \sin^2 \frac{1}{2}(\theta_i -\theta_j)} 
\end{equation}
where $\theta_j$ are the {\it eigenangles} in the three eigenvalues $e^{i\theta_j}$ of $u$ and $J$ is the Van de Monde determinant\footnote{Actually $J\equiv \sqrt{J^2}=\sqrt{D^*D}$, and $D=\prod^3_{i <  j } (e^{i\theta_i}-e^{i\theta_j})$ is Weyl's Van de Monde determinant\cite{Weyl}.}, the "Jacobian" of our parametrization \cite{Weyl}
\begin{equation}    \label{eq:J}
      J = \prod^3_{i <  j } 2 \sin\left(\frac{1}{2} (\theta_i- \theta_j)\right) .
\end{equation}
In mathematical terms $K_k$ and $M_k$ are off-toroidal derivatives which are non-commuting and may be represented by off-diagonal Gell-Mann matrices, see (\ref{eq:Kcomponents}) and (\ref{eq.Mcomponents}) below. The triple $K_k$ commute as body fixed angular momentum operators and $M_k$ "connect" the algebra by commuting into the subspace of $K_k$
\begin{equation}	\label{eq:KandMcommutation}
  [M_k,M_l]=[K_k,K_l]=-i\hbar K_m,
\end{equation}
cyclic in $k,l,m$. The components of ${\bf{K}}=(K_1,K_2,K_3)$ which are $SU(2)$-generators and ${\bf{M}}=(M_1,M_2,M_3)$ in the Laplacian carry spin and flavor. Interpreting $\bf K$ as an intrinsic spin operator is supported by the reversed sign in the commutator, like for body fixed coordinate systems in nuclear physics.

\begin{figure} 
\begin{center}
\includegraphics[width=0.45\textwidth]{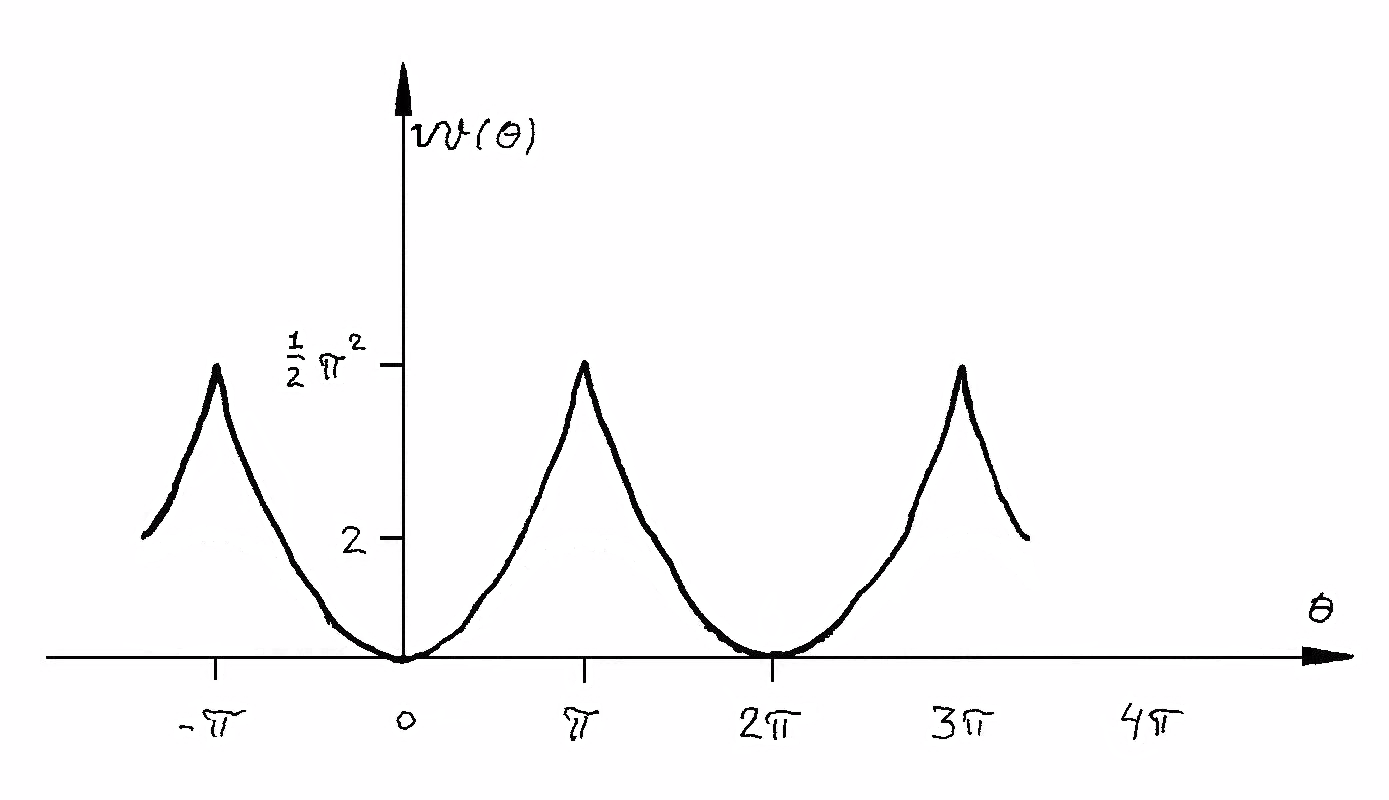}
\caption{Periodic parametric potential (\ref{eq:periodicpotential}) as a function of eigenangles of the $U(3)$ configuration variable.}
\label{fig:periodicpotential}
\end{center}
\end{figure}

The potential in (\ref{eq:schroedinger}) depends only on the eigenvalues of $u$ since the trace is invariant under conjugation $u\rightarrow vuv^{-1}$ by any $v\in U(3)$; in particular a conjugation that diagonalizes $u$. Thus
\begin{equation} \label{eq:mantonTracePotential}
  {\rm Tr}\ \chi^2\equiv d^2(e,u)=d^2(e,vuv^{-1})=d^2(v,vu).
\end{equation}
Here $e$ is the neutral element, the "origo" of $U(3)$. The last expression shows that the potential is left-invariant as are the coordinate fields that we shall soon introduce. In the above parametrization the potential reads
\begin{equation}
  \frac{1}{2} {\rm{Tr}}\ \chi^2\equiv W =w(\theta_1)+w(\theta_2)+w(\theta_3),
\end{equation}
i.e. a sum of periodic parametric potentials, see Fig. \ref{fig:periodicpotential}
\begin{equation} \label{eq:periodicpotential}
  w(\theta)     = \frac{1}{2}(\theta - n\cdot 2\pi)^2, \quad \theta \in [(2n-1)\pi,(2n+1)\pi], n\in \mathbb{Z}.
\end{equation}
The potential may be considered as the euclidean measure folded into the group manifold \cite{HansPlesnerJacobsen} in compliance with the space projection (\ref{eq:xprojection}) below.

Now, each of the nine generators $T_k$ of $U(3)$ implies directional derivatives locally at each point $u\in U(3)$ or so-called left-invariant coordinate fields
\begin{equation}    \label{eq:coordinateField} 
     \partial_k =  \frac{\partial}{\partial \alpha} ue^{i\alpha T_k} \vert _{\alpha=0} = u i T_k
\end{equation}
with related differential forms $d\alpha_k$, also called exterior derivatives $d\alpha_k(\partial_m)=\delta_{km}$. For the three toroidal degrees of freedom we use the angular symbols $\theta_j$. The quantization inherent in the basic equation (\ref{eq:schroedinger}) can then be expressed in a generalized action-angle form as
\begin{equation} \label{eq:commutationGeneralized}
  d\theta_i(\partial_j)=\delta_{ij} \Leftrightarrow [\partial_j,\theta_i]=\delta_{ij},
\end{equation}
where $d\theta_i$ are the torus forms and $\delta_{ij}$ is the Kronecker delta. By construction the act of the exterior derivative by a generator $X$ in the Lie algebra on a function $\Phi$ at a point $u$ in the Lie group manifold is
\begin{equation} \label{eq:exteriorderivativeDefinition}
  X_u[\Phi]\equiv d\Phi_u(X)=\frac{d}{dt}\Phi(ue^{tX})\!\mid_{t=0}.
\end{equation}
This was used\cite{TrinhammerEPL102} to generate the parton distributions in Fig. \ref{fig:partonDistribution}.

The three toroidal generators $T_j=-i\frac{\partial}{\partial\theta_j}=-i\partial_j\!\mid_e, j=1,2,3$ correspond to parametric momenta
\begin{equation}    \label{eq:toroidalParametricMomentum}
   p_j = - i \hbar \frac{1}{a} \frac{\partial}{\partial \theta_j} = \frac{\hbar}{a} T_j.
\end{equation}
and thus, corresponding to a space projection
\begin{equation}  \label{eq:xprojection}
   x_i=a\theta_i,
\end{equation}
we have the standard commutators
\begin{equation}
  [p_j,a\theta_i]=-i\hbar \delta_{ij}.
\end{equation}
In the above coordinate representation\cite{SchiffGellMannMatrices} the off-toroidal generators read \cite{TrinhammerEPL102}
\begin{align} \label{eq:Kcomponents}
   K_1& = a \theta_2p_3 - a \theta_3 p_2 = \hbar \lambda_7 \nonumber \\ 
     K_2& = a \theta_1p_3 - a \theta_3 p_1 = \hbar \lambda_5 \nonumber \\
     K_3& = a \theta_1p_2 - a \theta_2 p_1 = \hbar \lambda_2. 
\end{align}
and
\begin{align} \label{eq.Mcomponents}
  M_3/\hbar &= \theta_1 \theta_2 +  \frac{a^2}{{\hbar}^2} p_1 p_2 = \lambda_1 \nonumber \\
  M_2/\hbar &= \theta_3 \theta_1 +  \frac{a^2}{{\hbar}^2} p_3 p_1 = \lambda_4 \nonumber \\
  M_1/\hbar &= \theta_2 \theta_3 +  \frac{a^2}{{\hbar}^2} p_2 p_3 = \lambda_6 .
\end{align}
The lambdas are Gell-Mann generators \cite{SchiffGellMannMatrices}. From these and
\begin{gather} 
  Y/\hbar =
     \frac{1}{6} ( \theta^2_1 + \theta^2_2- 2\theta^2_3) 
   + \frac{1}{6} \frac{a^2}{{\hbar}^2}(p^2_1 + p^2_2- 2p^2_3) =\lambda_8/\sqrt{3}, \nonumber \\
  2I_3/\hbar =  \frac{1}{2} ( \theta_1^2 - \theta^2_2) + \frac{1}{2}  \frac{a^2}{{\hbar}^2}
   (p_1^2 - p^2_2) = \lambda_3
\end{gather}
the spectrum of $M^2$ was found to be \cite{TrinhammerEPL102}
\begin{align} 	\label{eq:M2spectrum}
  M^2 = \frac{4}{3}(n+\frac{3}{2})^2 - K(K+1) - 3  -\frac{1}{3}y^2 - 4i^2_3, \nonumber \\
  n = 0,1,2,3, \dots \end{align}
where $y$ and $i_3$ are hypercharge and isospin three-component quantum numbers. The minimum value for the positive definite $M^2$ is 13/4 in the case of spin 1/2, hypercharge 1 and isospin 1/2 as for the nucleon.

From here we are able to find specific solutions presented in  Secs. 3 and 4.

\section*{2.1. On flavor degrees of freedom}

The dynamics underlying the baryon mass spectroscopy in the present model is determined primarily by an intrinsic potential in the Lie group, namely the second term in (\ref{eq:schroedinger}). As for flavor degrees of freedom these are contained in the Laplacian on $U(3)$. The Laplacian can be parametrized in a polar decomposition with three toroidal, abelian derivatives and six off-torus derivatives (\ref{eq:Laplacian}). The latter correspond to the six off-diagonal Gell-Mann matrices (\ref{eq:Kcomponents}) and (\ref{eq.Mcomponents}). Three of these we interpret as spin generators and the remaining three are related to the isospin and hypercharge of the standard $SU_f(3)$ algebra (\ref{eq:M2spectrum}).

The last term in the Laplacian (\ref{eq:Laplacian}), the "centrifugal" term, can be integrated by exploiting the existence of the Haar measure over  $(\alpha_4,\alpha_5 \ldots,\alpha_9)$ by the factorization in (\ref{eq:factorizedPsi}) below. Further we use that the off-toroidal part of the wavefunction is an eigenstate of {\bf{K}}$^2$ and {\bf{M}}$^2$ together with the fact that the centrifugal term is symmetric under interchange of the torus angles $\theta_j$. 

The centrifugal term leads to a mass formula of the well-known Okubo type \cite{Okubo}. The spectrum of  {\bf{K}}$^2$ + {\bf{M}}$^2$ follows directly from (\ref{eq:M2spectrum}) to yield
\begin{equation}    \label{eq:K2M2spectrum}
  K(K+1) + M^2 = \frac{4}{3}\left(n+\frac{3}{2}\right)^2 - 3 - \frac{1}{3} y^2 - 4 i^2_3.
\end{equation}
It is natural in the present framework to classify the eigenstates according to the three independent values of $n$, $y$ and $i_3$. However we can make a transformation of this classification into the familiar one by rewriting the expression (\ref{eq:K2M2spectrum}) and choose the sum of $n$ and $y$ to be a constant. For $n + y =2$, which yields the lowest possible  $K(K+1) + M^2$, we get
\begin{equation}    
  K(K+1) + M^2 = \frac{40}{3}+(k^2_3 + m^2_3) - \frac{28}{3}y + 4\left[ \frac{1}{4}y^2\! - i(i+1)\right].
\end{equation}
Since $(K^2_3 + M^2_3)$  commutes with both $Y$ and $I^2$  we get for a given value of the quantum number $(k^2_3 + m^2_3)$
\begin{equation}    \label{eq:Okubo}
   K(K+1) + M^2 = a' + b' y +c'\left[\frac{1}{4}y^2 - i(i+1)\right]
\end{equation}
with the constants $a'=40/3+(k_3^2+m_3^2),\ b'=-28/3,\ c'=4$ respectively.

Equation (\ref{eq:Okubo}) is the famous Okubo mass formula that reproduces the Gell-Mann, Okubo, Ne'eman mass relations within the baryon N-octet and  $\Delta$-decuplet \cite{GellMann1962, Okubo, Neeman, Gasiorowicz} independently of the values of $a'$, $b'$ and $c'$. Of course this is only accurate if one chooses the same toroidal wavefunction for all members of a given multiplet. In practice the SU(3) symmetry breaking in (\ref{eq:Okubo}) will be influenced by the  $\mbox{\boldmath$\theta$}$-dependence in the centrifugal term because different values of {\bf{K}}$^2 +$ {\bf{M}}$^2$ lead to different values of the centrifugal potential and thereby influence which span of toroidal energy eigenstates will project out on a specific angular momentum eigenstate in the laboratory. Therefore, the $SU_f(3)$ symmetry break will not follow exactly (\ref{eq:Okubo}) in hypercharge.

\section*{2.2. On color degrees of freedom}

Hadronic phenomena are traditionally described in the standard model with interactions shaped by the gauge groups $SU_c(3)$ of their strong color interactions and $SU(2)\times U(1)$ of their electroweak interactions.

The model (\ref{eq:schroedinger}) uses the compact Lie group $U(3)$ as intrinsic configuration space. The maximal torus of $U(3)$ has three dimensions which we interpret as color degrees of freedom. We start off in a Hamiltonian framework with the Hamiltonian operating on states $\Psi(u)$, where $u$ is the configuration variable belonging to the Lie group $U(3)$. We can generate $SU(3)$-transforming color quark (\ref{eq:dPhi}) and gluon fields (\ref{eq:gluondPhi}) below from the exterior derivative on $\Psi$ scaled in measure by the Jacobian (\ref{eq:J}) of the polar decomposition. Summing over such color components (\ref{eq:dPhi}) for particular flavor tracks led to u and d valence quark parton distribution functions (Fig. \ref{fig:partonDistribution}) for an approximate protonic state via projections along mixed toroidal directions \cite{TrinhammerEPL102}. The parton distribution functions compare well with those established for the proton.

Because the dynamical structure is formulated on the Lie group, it will show different manifestations depending on which derivatives (\ref{eq:exteriorderivativeDefinition}) one is taking. For instance, the three toroidal dimensions for the color quark degrees of freedom are intermingled with flavor degrees of freedom since the hypercharge and isospin 3-component generators $Y$ and $I_3$ are not linearly independent of the three torodial generators $T_1, T_2, T_3$. And both are intermingled with the eight gluon dimensions laid out by the Gell-Mann matrices (\ref{eq:gluondPhi}) because these include generators proportional to $Y$ and $I_3$. Thus, we do not consider color and flavor degrees of freedom as being independent. For instance, the distribution functions in Fig. \ref{fig:partonDistribution} are produced by using the exterior derivative (\ref{eq:dPhi}) on tracks from the quark flavor generators\cite{TrinhammerEPL102} $T_u=2/3\ T_1-T_3$ and $T_d=-1/3\ T_1-T_3$. We actually see the reduction in the number of independent quark degrees of freedom as a reason that the baryon spectrum from (\ref{eq:schroedinger}) is not hampered by missing resonances as usual in ordinary quark models (QMs), compare Figs. \ref{fig:NDeltaSpectrum} and \ref{fig:missingResonances}.

\section*{3. Specific solutions of the model I. \\ Trigonometric base and the electron to neutron mass ratio}
\label{sec3}

It is possible to find the dimensionless eigenvalue for an unbroken neutron ground state ${\rm{E_n}}\equiv E_n/\Lambda$ in (\ref{eq:schroedinger}) with quite high precision by a Rayleigh-Ritz method\cite{Bruun}. We factorize the wavefuction $\Psi$ in (\ref{eq:schroedinger}) into a toroidal part $\tau$ and an off torus part $\Upsilon$
\begin{equation}	\label{eq:factorizedPsi}
  \Psi(u)=\tau(\theta_1,\theta_2,\theta_3)\Upsilon(\alpha_4,\alpha_5,\alpha_6,\alpha_7,\alpha_8,\alpha_9).
\end{equation}
In that way (\ref{eq:schroedinger}) can be solved for specific choices of spin and flavor inflicted by the six off torus generators contained in the Laplacian. After integration over the six off-toroidal degrees of freedom $\alpha_4,\alpha_5,\alpha_6,\alpha_7,\alpha_8,\alpha_9$ one ends up with a Schr\"{o}dinger equation
\begin{equation}    \label{eq:toroidalSchroedinger}
 [ -\frac{1}{2}\sum^3_{j=1} \frac{\partial^2}{\partial \theta_j^2}+V]R(\theta_1,\theta_2,\theta_3) = \mbox{E}R(\theta_1,\theta_2,\theta_3).
\end{equation}
Here $R=J\tau$ with $J$ from (\ref{eq:J}) and
\begin{gather}    
  V = -1 + \frac{1}{2}\cdot\frac{4}{3}\sum^3_{ i <  j} \frac{1}{8 \sin^2 \frac{1}{2}(\theta_i -\theta_j)} \\ \nonumber
 +  w(\theta_1) + w(\theta_2)+ w(\theta_3).
\end{gather}
contains in the second term contributions from off-toroidal degrees of freedom that carry spin and flavor in the specific choice here of spin, hypercharge and isospin $s=1/2, y=1, i=1/2$. The numerator $4$ in front of the sum is the minimum value of $({\bf K}^2 + {\bf M}^2)/\hbar^2$ for this combination as well as for $s=3/2, y=1, i=3/2$ corresponding to the choices respectively of $n=1$ and $n=2$ in (\ref{eq:M2spectrum}). The term correponds to the centrifugal potential when solving the hydrogen atom in polar coordinates (\ref{eq:radialLaplacian}). The constant term is a global curvature term \cite{Dowker} arising from differentiating through $J^2$ in the Laplacian (\ref{eq:Laplacian}).

\begin{figure}
\begin{center}
\includegraphics[width=0.45\textwidth]{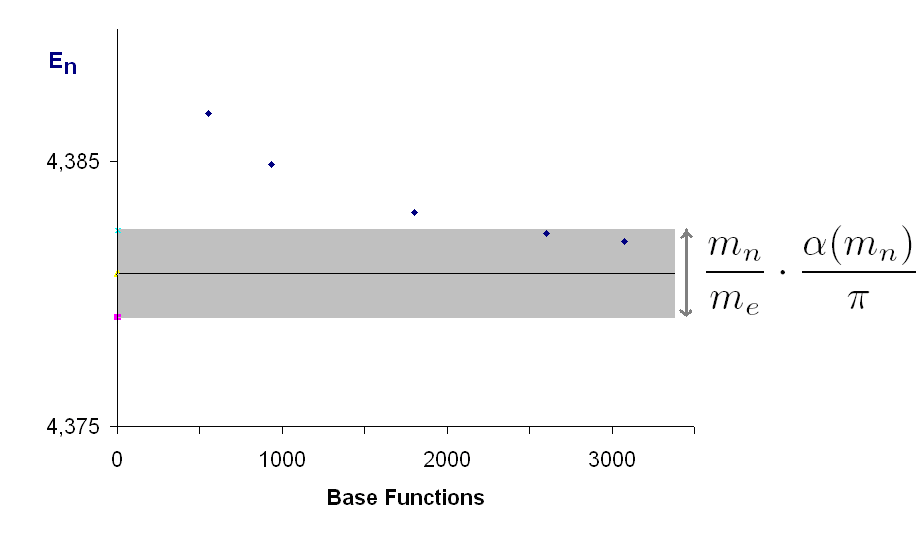}
\caption{The ground state eigenvalue \cite{TrinhammerEPL102} $\rm E_n$ (dots) from (\ref{eq:schroedinger}) compared with the expected result from the neutron to electron mass ratio with a sliding scale\cite{LandauLifshitzVol4p603,RPP2012p136,WeinbergQToFIIp158and126} fine structure constant $\alpha(m_n)$ (grey band). The grey band shows the incertainty in the estimate for $\alpha(m_n)$ at nucleonic energies.}
\label{fig:memnConvergence}
\end{center}
\end{figure}

The measure-scaled toroidal wavefunction $R$ can be expanded on solutions $b$ to the separable problem
\begin{equation}    \label{eq:SeperableSchroedinger}
 [ -\frac{1}{2}\sum^3_{j=1} \frac{\partial^2}{\partial \theta_j^2}+W]b(\theta_1,\theta_2,\theta_3) = \mbox{E}b(\theta_1,\theta_2,\theta_3).
\end{equation}
Due to the arbitrary labeling of the eigenangles $\theta_j$, the toroidal wavefunction $\tau$ is symmetric in these and as $J$ is antisymmetric so must be $R=J\tau$. Therefore solutions to (\ref{eq:toroidalSchroedinger}) and (\ref{eq:SeperableSchroedinger}) can be constructed from Slater determinants \cite{Slater}
\begin{equation}
  b_{pqr}=\epsilon_{ijk}b_p(\theta_i)b_q(\theta_j)b_r(\theta_k),
\end{equation}
where $p,q,r$ are natural number labels for orthogonal solutions to the one-dimensional Sch\"{o}dinger equation
\begin{equation} \label{eq:onedimSchoedinger}
  [ -\frac{1}{2}\frac{\partial^2}{\partial \theta^2}+w(\theta)]b_p(\theta)=e_pb_p(\theta)
\end{equation}
with periodic parametric potential. This is postponed to Sec. 4. Here we will use an expansion set where the necessary integrals for the Rayleigh-Ritz procedure can be found analytically. The measure scaled toroidal part $R$ of the wavefunction is expanded on "trigonometric" Slater determinants
\begin{equation} \label{eq:Slater}
  f_{pqr}(\theta_1,\theta_2,\theta_3)=\epsilon_{ijk}\cos p\theta_i
\sin q\theta_j \cos r\theta_k,
\end{equation}
where $p,q,r$ are integers, $p=0,1,2,...,P-1$; $q=1,2,3,...,P$; $r=p+1,p+2,...,P$. The order parameter $P$ determines the number of independent states on which we expand. The value $P=12$ corresponds to 936 expansion functions and yields ${\rm{E_n}}=4.3849$ whereas $P=18$ with 3078 expansion functions yields ${\rm{E_n}}=4.3820$ seen in Fig. \ref{fig:memnConvergence}. Calculations of ${\rm{E_n}}$ with higher values of $P$ are beyond the handling capacity of our computer programmes \cite{BaryonLieProgrammes}.

\section*{4. Specific solutions of the model II.\\ Parametric base and the baryon spectrum} 
\label{sec4}

\begin{figure} 
\begin{center}
\includegraphics[width=0.45\textwidth]{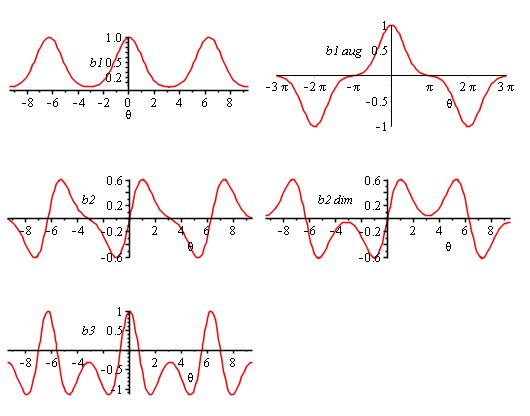}
\caption{Parametric eigenfunctions from (\ref{eq:onedimSchoedinger}). The period doubling (right) in the diminished state for level two is paired with an augmented period doubled state for level one (above).}
\label{fig:parametric}
\end{center}
\end{figure}
\begin{figure}
\begin{center}
\includegraphics[width=0.45\textwidth]{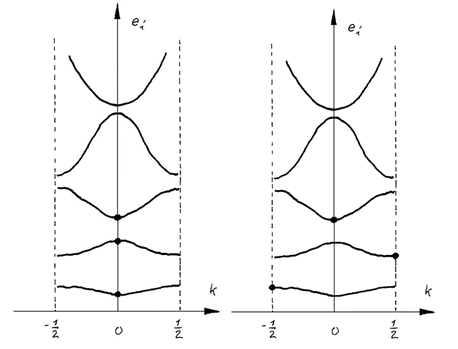}
\caption{Reduced zone scheme \cite{AshcroftMermin} for parametric eigenvalues. The black dots represent the values for the unstable neutron state (left) and the proton state (right). For clarity the variation of the eigenvalues with Bloch wave number $\kappa$ is grossly exaggerated for the lowest states.}
\label{fig:reducedzone}
\end{center}
\end{figure}

\begin{figure}
\begin{center}
\includegraphics[width=0.45\textwidth]{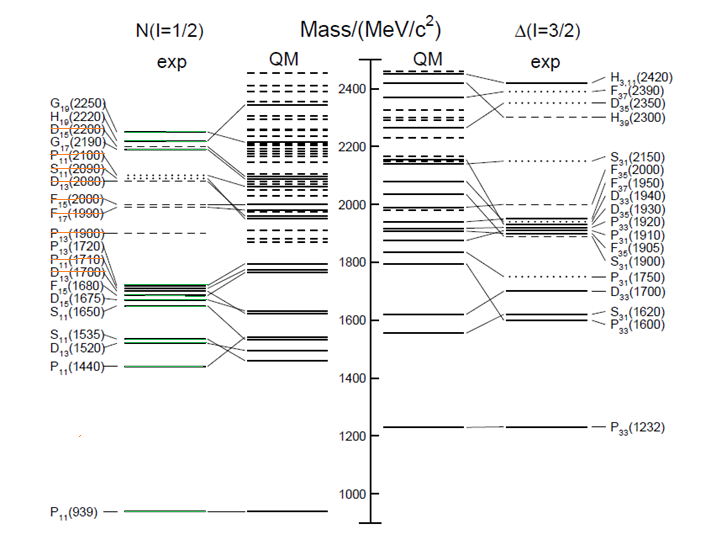}
\caption{The missing resonance problem of baryon spectroscopy. Figure adapted from Review of Particle  Physics \cite{RPP2012p205}. Too many resonances are predicted from ordinary quark models (QM) than are experimentally observed (exp). Four star $N$ resonances are highlighted by green whereas three star resonances are crossed through in orange. Compare with the neutral flavor spectrum in Fig. \ref{fig:NDeltaSpectrum} predicted from the model (\ref{eq:schroedinger}) where the number of predicted states match the observed four star resonances.}
\label{fig:missingResonances}
\end{center}
\end{figure}

We now consider the expansion on Slater determinants constructed from solutions to (\ref{eq:onedimSchoedinger}). Figure \ref{fig:parametric} shows solutions for the first three eigenvalues $e_1, e_2, e_3$. The structures of (\ref{eq:SeperableSchroedinger}) and (\ref{eq:toroidalSchroedinger}) with periodic potentials either $V$ or $W$ imply the introduction of Bloch wave expansion factors
\begin{equation} \label{eq:gp}
  g_p(\theta)=e^{i\kappa\theta}u_p(\theta),
\end{equation}
where $\kappa$ introduces the Bloch degree of freedom. We shall argue that the Bloch degrees of freedom are opened by a Higgs mechanism that will allow a diminishing of the ground state eigenvalue via the creation of the $\nu_e, e_L$ doublet and it's coupling to a Higgs field. For instance the ground state eigenvalue $\rm{E_n}=e_1+e_2+e_3=4.47...$ of (\ref{eq:SeperableSchroedinger}) is lowered to a value $\rm{E_p}=e'_1+e'_2+e_3=4.46...$ for real symmetry broken states of parametric eigenvalues $e'_1$ and $e'_2$ with $4\pi$ periodicity analogous to $\kappa_1,\kappa_2=\pm\frac{1}{2},\pm\frac{1}{2}$ for the Bloch-phase-containing $g$s as opposed to the $2\pi$ periodicity of the $b_p$s and $u_p$s, see Fig. \ref{fig:reducedzone}. The period doublings are allowed since they leave the square of the wave function $\Psi^2$ singlevalued on $U(3)$. Table \ref{tab:paraeig} shows results for the parametric eigenvalues. Different combinations of three different levels give a good reproduction of the observed spectrum of all the certain (four star) neutral flavor baryon resonances, i.e. the N and $\Delta$ spectrum with {\it no missing resonance problem} \cite{RPP2012p205}, compare Figs. \ref{fig:NDeltaSpectrum} and \ref{fig:missingResonances}. 

By summing up the three lowest levels we get an approximate estimate of the relative neutron to proton mass shift
\begin{equation}
  \frac{m_n-m_p}{m_p}\approx\frac{e_1+e_2+e_3-(e'_1+e'_2+e_3)}{e'_1+e'_2+e_3}=0.13847{\%}.
\end{equation}
This is to be compared with the value $0.137842{\%}$ calculated from the observed neutron and proton masses which are known experimentally with eight significant digits \cite{RPP2012p79}. The exact value for $\rm{E_n}$ from (\ref{eq:toroidalSchroedinger}) is $4.38...$ which is a few percent lower than the approximate value $\rm{E_n}=4.47...$ mentioned above. A suitable base on which to expand an exact calculation for $\rm{E_p}$ has not been found.

In Fig. \ref{fig:NDeltaSpectrum} we use the scale $\Lambda\equiv\hbar c/a$ for the approximate solutions from the proton rest energy $\Lambda=E/\rm{E}=$938.3 MeV/4.468 = 210 MeV. The predicted spectrum agrees with the number and grouping of all the certain resonances in the ${\rm{N}}\Delta$-sector. By 'certain' we mean all the well established resonances with four stars in the particle data group listings \cite{RPP2012}. 

Only one observed certain N-resonance in the group of three resonances in the domain around 1500 MeV is missing in the predictions. However the approximate treatment in (\ref{eq:SeperableSchroedinger}) suggests a neutral singlet 1,3,5 at 1510 MeV exactly in that area. The exact treatment case in (\ref{eq:toroidalSchroedinger}) can be solved for neutral states by a Rayleigh-Ritz method which places the singlet at 1526 MeV, see Table \ref{tab:singlets}. This state is thought to mix with the other two $\rm{N}$-resonances nearby to give the total of three $\rm{N}$-resonances in the group. The next singlet state 1,3,7 is predicted in the "desert" area between 1700 MeV and 2100 MeV. In the approximate case the resonance comes out at 1965 MeV and in the exact case it comes out at 2051 MeV (Table \ref{tab:singlets}). No certain N-resonance is observed in this domain. Being close to the observed resonance domain around 2100-2200 MeV the state 1,3,7 might hide itself by mixing with ordinary $\rm{N}$-states. On the other hand it could explain in particular the neutral charge manifestation of a new resonance N(2040) seen in $m_{p\pi^-}$ invariant mass spectra\cite{KlemptRichard} from $J/\Psi\rightarrow p\pi^-\bar{n}$. The similar electrically neutral singlet 5,7,9 at 4499 MeV lies just above the free charm threshold at 4324 MeV for baryonic decay into $\Sigma_c^+(2455)D^-$ and should be visible (together with 3,5,11 at 4652 MeV and 1,7,11 at 4723 MeV) in neutron diffraction dissociation experiments like those in reference \cite{Aleev}. They should all be visible in $\pi^-p\to\pi^-p$ scattering like in ref.\cite{Epecur} and in $\gamma n\rightarrow p\pi^-$ photoproduction experiments like in ref. \cite{Zhu}. The resonance 1,7,11 at 4723 MeV is expected to be particularly pronounced since it contains level 1 which lies as the deepest in the geodetic potential wells. Other, lower lying, neutral electric charge, neutral flavor singlets shown in Table \ref{tab:singlets} might be visible in $m_{p\pi^-}$ invariant mass  from $B$ decay experiments like in ref. \cite{Babar}. Note that the neutral flavor singlets have no electrically charged partners. This distinguishes the model (\ref{eq:schroedinger}) predictions from standard quark flavor models.

\begin{table}
\begin{center}
\caption{Parametric eigenvalues (\ref{eq:onedimSchoedinger}) to construct the approximate baryon spectrum in Fig. \ref{fig:NDeltaSpectrum}. The eigenvalues are calculated with 1500 collocation points. The lowest eigenvalues, as expected, are close to those of the ordinary harmonic oscillator. Moving up to higher levels the eigenvalues differ more and more from those of the harmonic oscillator as indicated in Fig. \ref{fig:reducedzone}. The lower levels have been calculated by three different methods with discrepancies only from the eights significant digit, see Table \ref{tab:numericalmethods}. \vspace{3mm}}
\label{tab:paraeig}
\begin{tabular}{ c c c c } \hline\hline\\
$p$&$e_p$&$e'_p$&$e'_p$ \\
Level&Eigenvalue&Diminished&Augmented \\ \hline \\
1&0.499804708&&0.5001727904 \\
2&1.502988968&1.496433950& \\ 
3&2.471378779&&2.522629649 \\
4&3.600509000&3.377236032& \\
5&4.218515963&&4.803947527 \\
6&6.197629004&5.160535373& \\
7&6.383117406&&7.820486992 \\
8&9.688466291&7.922699154& \\ 
9&9.751335596&&11.80644676 \\
10&14.1755275&11.84897047& \\
11&14.2063708&&16.79575229 \\ \hline\hline
\end{tabular}
\end{center}
\end{table}

\begin{table}
\begin{center}
\caption{Comparison of numerical results for the eigenvalue of the ground state. The seperable problem (\ref{eq:SeperableSchroedinger}) has been solved by four different methods three of which gives a set of eigenvalues for the one-dimensional problem (\ref{eq:onedimSchoedinger}) from which the eigenvalues for the three-dimensional problem (\ref{eq:SeperableSchroedinger}) is constructed. These eigenvalues can be used to check the Rayleigh-Ritz method for solving the three-dimensional problem directly. Mutual discrepancies are due to the finite expansions in the different methods. The fine agreement among the different methods \cite{BaryonLieProgrammes} lends support to the Rayleigh-Ritz method also for solving the full eq. (\ref{eq:toroidalSchroedinger}).\vspace{3mm}}
\label{tab:numericalmethods}
\begin{tabular}{c c c c} \hline\hline \\
1D-level & Iterative & MacLaurin & Collocation\\
number & integration & series \cite{PovlHolm} & 1500\\
$p$& & & points\\
& Comal & - & Matlab $m_n$ \\ \hline\\
1&0.499804708&0.499804704&0.499804708 \\ 
2&1.502988981&1.502988968&1.502988968 \\
3&2.471378882&2.471378899&2.471378779 \\ 
Sum&4.474172571&4.474172571&4.474172455 \\ \hline\hline
\end{tabular}
\end{center}
\end{table}

\begin{table}
\begin{center}
\caption{Scarce singlet states. Eigenvalues based on Slater determinants of three cosines up to order 20 analogous to (\ref{eq:Slater}). The first column shows eigenvalues of the approximate equation (\ref{eq:SeperableSchroedinger}) and the third column shows eigenvalues of the exact equation (\ref{eq:toroidalSchroedinger}). A singlet 579-like resonance is predicted at 4499 MeV in the free charm system $\Sigma_c^+(2455)D^-$  slightly above its threshold at 4324 MeV. The rest masses are predicted from a common fit of the nucleon ground state 939.6 MeV to the ground state 4.38 of (\ref{eq:schroedinger}) resp. (\ref{eq:toroidalSchroedinger}) with no period doublings. \vspace{3mm}}
\label{tab:singlets}
\begin{tabular}{c c c c c} \hline\hline \\
Singlet & Toroidal & Singlet & Rest mass\\
approximate (\ref{eq:SeperableSchroedinger}) & label & exact (\ref{eq:toroidalSchroedinger}) & MeV/c$^2$\\ \hline \\
7.1895&1 3 5&7.1217&1526 \\ 
9.3568&1 3 7&9.5710&2051 \\ 
11.1192&1 5 7&11.2940&2420 \\ 
12.7175&1 3 9&13.2505&2839 \\ 
13.0927&3 5 7&13.2811&2846 \\ 
14.4494&1 5 9&14.9641&3206 \\ 
16.4086&3 5 9&16.9213&3626 \\ 
16.6605&1 7 9&17.3006&3707 \\
17.1769&1 3 11&18.0090&3859 \\
18.6320&3 7 9&19.2577&4126 \\
18.9214&1 5 11&19.7327&4228 \\
20.3774&5 7 9&20.9940&4499 \\
20.8910&3 5 11&21.7110&4652 \\
21.0766&1 7 11&22.0409&4723 \\ \hline\hline
\end{tabular}
\end{center}
\end{table}

\section*{5. Where are the quarks? - Projection to space}
\label{sec5}

\begin{figure} 
\begin{center}
\includegraphics[width=0.45\textwidth]{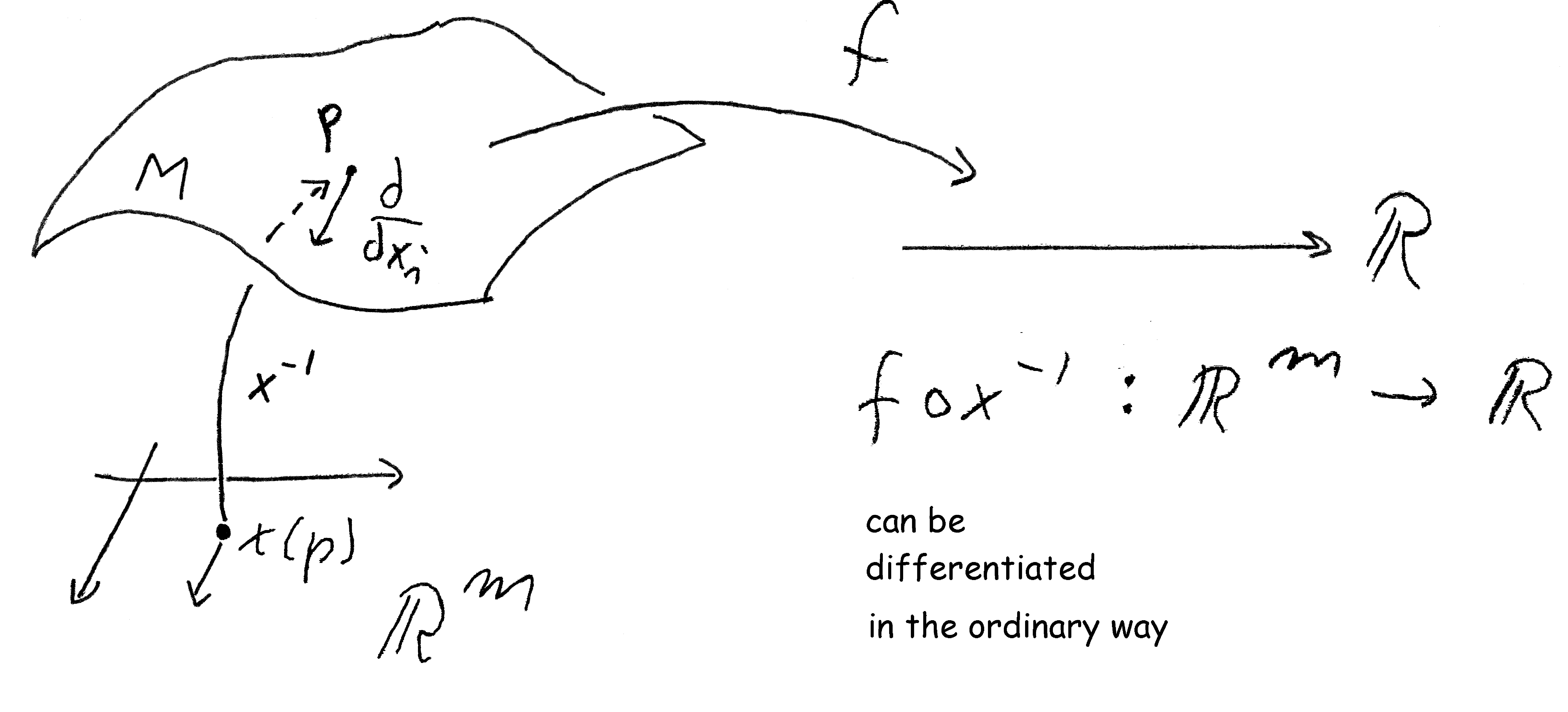}\caption{Derivation of a real-valued function $f$ at point $p$ in the manifold $M$ is defined by using a local smooth map $x:M\rightarrow \mathbb{R}^m$ to pull back the problem to an ordinary derivation on $\mathbb{R}^m$ by using the pullback function  $f\circ x^{-1}:\mathbb{R}^m\rightarrow \mathbb{R}$. One can then differentiate $f\circ x^{-1}$  in the ordinary way. This idea is readily generalized to a complex-valued function and in the present case the manifold $M$ could be $U(3)$ and the then complex-valued function $f$ could be either the wavefunction $\Psi$ or its measure-scaled partner  $\Phi$.}
\label{fig:projectionParameterSpace}
\end{center}
\end{figure}

Here we look at projections of the wavefunction to fields in laboratory space. For each element  $u\in U(3)$  we have a corresponding left-translation $l_u$  on  $v \in U(3)$
\begin{equation} 
   l_u(v)\equiv uv,
\end{equation}
and for any left-invariant vector field  X we have \cite{CornwellAppendixJ}
\begin{equation} 
   \mbox{X}_{uv} =d(l_u)_v(\mbox{X}_v).
\end{equation}
In particular we have for the toroidal coordinate fields when comparing with (\ref{eq:coordinateField})
\begin{equation} \label{eq:leftinvariantDerivative}
   \partial_j \!\mid_{u\cdot e} = d(l_u)_e(\partial_j \!\mid_e) = u \partial_j \!\mid_e.
\end{equation}
Thus the exterior derivative $d$  acts as the identity on left-translations at the origo $e$, i.e. the algebra approximates the group in the vicinity of origo. We now expand the exterior derivative, also called the momentum form \cite{HolgerBechNielsen}, of the measure-scaled toroidal wave function $R = J\tau$  on the torus forms (\ref{eq:commutationGeneralized}), i.e.  
\begin{equation} \label{eq:dPhi}
   dR= \psi_j d \theta_j,
\end{equation}
where the coefficients are the local partial derivatives \cite{GuilleminPollack}
\begin{equation} 	\label{eq:quarkdR}
    \psi_j(u)\equiv dR_u (\partial_j), \quad j=1,2,3.
\end{equation}
For the coefficients we have by left-invariance (\ref{eq:leftinvariantDerivative}) 
\begin{gather} 
   \psi_j(u)=dR_u(\partial_j)= \partial_j \!\!\mid_u[R] =u\partial_j \!\!\mid_e[R]\\ \nonumber
    = ud R_e(\partial_j) = u \psi_j(e). \hspace{1cm} \end{gather}
The sum of the differential components of the torus form will inherit the left-invariance
\begin{gather}  \label{eq:psioverline}
   \overline{\psi}(u) \equiv \psi_1(u) + \psi_2(u) + \psi_3(u)   =\\ \nonumber
 u (\psi_1(e) + \psi_2(e) + \psi_3(e)) = u \overline{\psi} (e). 
\end{gather}
Now, in particular $\psi_j(e)=dR_e(\partial_j)=\partial R/\partial\theta_j$ belongs to the tangent space $TM_e$ of the maximal torus $M$ at $e$ and therefore so does their sum $\overline{\psi}(e)$ 
as in general $\psi_j(u)\in TM_u$. The set of generators  $\{ i T_j \}$ are the coordinate field generators $\partial_j$ which also constitute an induced base from parameter space
\begin{equation} 
  \partial_j \!\! \mid_u = \frac{\partial}{\partial\theta_j} \!\! \mid_u = d(exp)_{exp^{-1}(u)}(\vec{c_j}),
\end{equation}
where $\{\vec{c}_j \}$ is a set of base vectors for the parameter space for the torus, see Fig. \ref{fig:projectionParameterSpace}. 

\begin{figure}
\begin{center}
\includegraphics[width=0.45\textwidth]{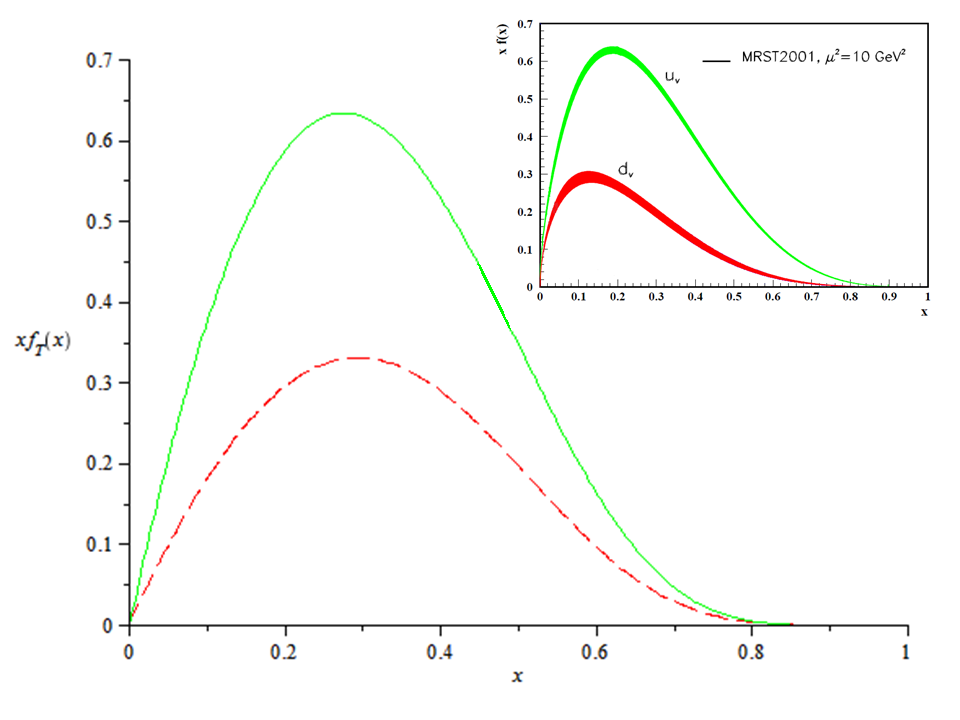}
\caption{Valence quark parton distribution functions \cite{TrinhammerEPL102} for u quarks (solid, green) and d quarks (dashed, red) for an approximate protonic state compared with established results adapted from the Particle Data Group \cite{RPP2004} (insert) with other parton distributions erased.}
\label{fig:partonDistribution}
\end{center}
\end{figure}

In our interpretation we identify $\{\vec{c}_j \}$ as a base for a fundamental representation space for the color algebra $su(3)$ at a particular point $P(x,y,z)$ in laboratory space. We may thus introduce at $P$ complex-valued components $\tilde{\psi_j}$ for the color vector $\vec{\psi}$ and write
\begin{equation} \label{eq:psiVector}
   \vec{\psi} = \tilde{\psi}_1 \vec{c}_1 + \tilde{\psi}_2 \vec{c}_2 + \tilde{\psi}_3 \vec{c}_3 =
\begin{Bmatrix}  \tilde{\psi}_1 \\ \tilde{\psi}_2 \\ \tilde{\psi}_3 \end{Bmatrix}.
\end{equation}
In the above representation $u$ will be represented by a $3\times 3$ matrix $U$. For rotations under $V\in SU(3)$ at $P$ we then have
\begin{equation} \label{eq:cRotated}
   \vec{c}_j \rightarrow \vec{c'}_j = V\vec{c}_j 
\end{equation}
and
\begin{equation} \label{eq:Urotated}
   U \rightarrow U'=VUV^{-1}.
\end{equation}
From (\ref{eq:cRotated}) and (\ref{eq:Urotated}) we can derive the transformation property of $\overline{\psi}(u)$.
\begin{gather} 
   \overline{\psi}(u) = U \overline{\psi}(e) \rightarrow \overline{\psi}(u') ' = U' \overline{\psi}(e) ' \\ \nonumber =
    VUV^{-1}V \overline{\psi}(e) = VU \overline{\psi}(e) = V \overline{\psi} (u),
\end{gather}
which shows that the differential component vector $\overline{\psi}$ transforms as a color vector in the fundamental representation under $SU(3)$ rotations. In other words left-translation in group space projects out as $SU(3)$ rotation in projection space. We thus interpret $\overline{\psi}$ as a quark field with three color components which may be projected on a specific base like in (\ref{eq:psiVector}). 
The distributions\cite{TrinhammerEPL102} in Fig. \ref{fig:partonDistribution} are for $T_u=\frac{2}{3}T_1-T_3$ and $T_d=-\frac{1}{3}T_1-T_3$.

Likewise the gluon fields may be seen as resulting from a projection on adjoint representation spaces of an expansion of the momentum form corresponding to the full set of eight generators $\lambda_k$  needed to parametrize the general group element $u=e^{i\alpha_k\lambda_k}$  -- separating out a phase factor \cite{BrodskyEtAlPhaseFactor}. Thus for each generator $T_a$  we have left-invariant vector fields $\partial_a$  defined as
\begin{equation} 
   \partial_a = \frac{\partial}{\partial \omega} e^{i\alpha_k T_k} e^{i \omega T_a} \!\! \mid_{\omega=0}
   = u i T_a,
\end{equation}
where $T_a=-i\partial/\partial \alpha_a = - i \partial_a\! \! \mid_e$. We now choose the set $\{T_a\}$  as a base for the adjoint representation. This base transforms under $V \in SU(3)$  like
\begin{equation} 
   T_a' = VT_aV^{-1}.
\end{equation}
Analogous to (\ref{eq:dPhi}) we expand the exterior derivative $d\Phi$  of the full measure-scaled wavefunction $\Phi=J\Psi$ on forms related to the left-invariant vector fields $\partial_a$ to get the adjoint projection field
\begin{equation} \label{eq:gluondPhi}
   \overline{A}(u) = \sum_a d \Phi_u(\partial_a).
\end{equation}
We want to show that $\overline{A}$  transforms according to the adjoint representation. First we have the equivalent of (\ref{eq:psioverline})
\begin{gather} 
   \overline{A}(u) = \sum_a d \Phi_u(\partial_a)
      =\sum_a \partial_a \!\! \mid_u [\Phi]
      = \sum_a u \partial_a \!\! \mid_e [\Phi]
      \\ \nonumber =u \sum_a d\Phi_e(\partial_a)=u\overline{A}(e).
\end{gather}
Here we understand in analogy with (\ref{eq:psiVector}) that
\begin{equation}
	\overline{A}(e)= \tilde{A}_a(e)T_a,
\end{equation}
where again $\tilde{A}_a$ are complex-valued components and $\{T_a\}$ was the adjoint base.
We may then proceed to show the adjoint transformation property of $\overline{A}$
\begin{gather}
	\overline{A}(u) \rightarrow \overline{A}(u') ' = U'\overline{A}(e) ' = U'\tilde{A_a}(e)T_a'
	\nonumber \\ =VUV^{-1} \tilde{A_a}(e)VT_aV^{-1} =VUV^{-1}V \tilde{A_a}(e)T_aV^{-1} 
	\nonumber \\
	= VU \overline{A}(e)V^{-1} = V \overline{A}(u)V^{-1},
\end{gather}
which corresponds to the gauge group rotation transformation property of the gluon fields\cite{DonoghueEtAlGaugeTransformation} $B$ 
\begin{equation} \label{eq:BgaugeTrans}
	B_{\mu} ' = VB_{\mu}V^{-1}+\frac{i}{g}(\partial_{\mu}V)V^{-1},
\end{equation}
where
\begin{equation}
	V = e^{-i \alpha_a(x) T_a}.
\end{equation}
We note that as space time fields the gauge fields also acquire a term representing the variation along spacetime translations as represented by the second term in (\ref{eq:BgaugeTrans}). Note further that translational invariance in group space corresponds to an $SU(3)$ rotational invariance in representation space and thereby the translational invariance of the interaction potential (\ref{eq:mantonTracePotential}) in group space through the projections (\ref{eq:xprojection}) and (\ref{eq:dPhi}) reflects the gauge invariance of the fields in laboratory space.

\section*{6. An exemplar Higgs mechanism for the neutron decay}
\label{sec6}

We want to relate the strong interaction dynamics inherent in (\ref{eq:schroedinger}) to the electroweak interaction involved in e.g. the decay of the neutron to the proton. We settle the relation through the structure of the geodetic potential (\ref{eq:periodicpotential}) supplemented by a {\it trailing} ansatz ($\hbar=c=1)$
\begin{equation}	\label{eq:HiggsAnsatz}
 \Lambda\theta=\alpha\phi
\end{equation}
which balances the products of the coupling constants and the corresponding ("phase") fields for the respective interactions, namely a color field $\theta$ and a higgs field $\phi$. The color field is a space projection of an eigenangle dynamical variable from the description of the intrinsic configuration space and as such starts off dimensionless whereas $\phi$ - or more precisely $\sqrt{\hbar c}\phi$ - has the dimension of energy \cite{GriffithsElementaryParticlesp403}. A certain caution is therefore needed when the geodetic potential for $\theta$ is to be translated into a potential for $\phi$.

Let us first look at the Klein-Gordon Lagrangian\cite{GriffithsElementaryParticlesp355} for a scalar field $\phi$ of mass $m$ 
\begin{equation}
  L=\frac{1}{2}\partial_\mu\phi\partial^\mu\phi-\frac{1}{2}(\frac{mc^2}{\hbar c})^2\phi^2.
\end{equation}
While in the following we set $\hbar=c=1$, we still want to keep track of the length scale and we therefore write
\begin{equation}
  L=\frac{1}{2}\partial_\mu\phi\partial^\mu\phi-\frac{1}{2}(\frac{\rm m}{\tilde{a}})^2\phi^2
\end{equation}
Here $\rm m$ is dimensionless while $m=\rm m/\tilde{a}$ represents the mass of dimension $\rm L^{-1}$ in usual $\hbar=c=1$ notation, thus the well known expression
\begin{equation}	\label{eq:LagrangianScalarField}
  L=\frac{1}{2}\partial_\mu\phi\partial^\mu\phi-\frac{1}{2}m^2\phi^2.
\end{equation}
Corresponding to the length scale $\tilde{a}$ we have an energy scale $\tilde{\Lambda}=\hbar c/\tilde{a}$ which we shall settle below in (\ref{eq:tildeenergyscale}). The dimensionless mass $\rm m$ is also given as ${\rm m}=mc^2/\tilde{\Lambda}$

\begin{figure}
\begin{center}
\includegraphics[width=0.45\textwidth]{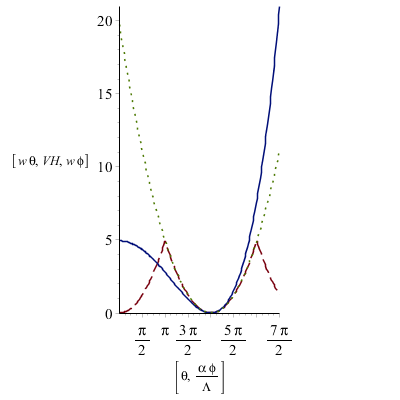}
\caption{A neighbouring trough in the periodic intrinsic potential (dashed, red) fitted by a lifted Higgs potential (solid, blue). In dotted green is shown the restriction (\ref{eq:periodicphipotential}) around the shifted minimum of the geodetic potential. All three curves exhibit a harmonic form for small pertubations with the same second order mass term used to derive the Higgs mass (\ref{eq:Higgsmass}) and share the shift to derive the electroweak energy scale (\ref{eq:v}).}
\label{fig:higgspotfit}
\end{center}
\end{figure}

In the Higgs mechanism the mass term for a Higgs particle field $H$ equivalent to the mass term in  (\ref{eq:LagrangianScalarField}) follows from the second order derivative of a Higgs potential $V_H(\phi)$ at a minimum point $\phi_0\neq 0$. The geodetic potential (\ref{fig:periodicpotential}) has such minimum points off 0, which are "activated" in the neutron decay when parametric period doublings occur. The period doublings correspond to sudden jumps of $\theta$ from one trough of the geodetic potential to a neighbouring one. We therefore consider a match of a Higgs potential to a dimensionful edition
\begin{equation}	\label{eq:periodicphipotential}
  w(\theta)\to\frac{1}{2}(\phi-\phi_0)^2
\end{equation}
of the geodetic potential (\ref{eq:periodicpotential}) neighbouring to the generic $n=0$ section. With the balancing trailing ansatz (\ref{eq:HiggsAnsatz}) we find that a jump from $\theta=0$ to $\theta=2\pi$ corresponds to
\begin{equation}	\label{eq:tildeenergyscale}
  \phi_0=2\pi\frac{\Lambda}{\alpha}\equiv\tilde{\Lambda}.
\end{equation}
To match (\ref{eq:periodicphipotential}) we introduce a constant term $\delta^2$ in the exemplar Higgs "potential" \cite{WeinbergQToFIIp303,GriffithsElementaryParticlesp381} to have
\begin{equation}	\label{eq:liftedhiggspotential}
  V_H(\phi)=\delta^2-\frac{1}{2}\mu^2\phi^2+\frac{1}{4}\lambda^2\phi^4,  \ \ \phi^2=\phi^\dagger\phi.
\end{equation}
This potential as usual has minima at $\phi_0^2=\mu^2/\lambda^2$. Note, that the essential  thing for the Englert, Brout, Higgs, Guralnik, Hagen, Kibble-mechanism \cite{EnglertBrout,HiggsSep1964,HiggsOct1964,GuralnikHagenKibble}  is not the particular shape\cite{Higgs1966} of $V_H$ but the fact that $V_H$ has a minimum for non-zero $\phi$. For a real scalar $\phi$ we find (\ref{eq:liftedhiggspotential}) to match (\ref{eq:periodicphipotential}) in the neighbourhood of $\phi_0$ for
\begin{equation}
  \delta^2=\frac{1}{8}\phi_0^2,\ \mu^2=\frac{1}{2},\ \lambda^2=\frac{1}{2}\frac{1}{\phi_0^2},
\end{equation}
see Fig. \ref{fig:higgspotfit}. With these choices $w$ in (\ref{eq:periodicphipotential}) and $V_H$ in (\ref{eq:liftedhiggspotential}) both agree at $\phi_0$ up to second order, where $V_H''(\phi_0)=2\mu^2$.
For a complex $\phi=(\phi_1(x)+i\phi_2(x))/\sqrt{2}$ we have minima in (\ref{eq:liftedhiggspotential}) for
\begin{equation}
  \phi=\phi_0e^{i\beta},
\end{equation}
where $\beta$ is a real phase angle and for convenience we define $v=\sqrt{2}\phi_0\equiv \tilde{v}\phi_0$. Since
\begin{gather}
  V_H(\phi_0+H/\sqrt{2})
  =-\frac{1}{2}\mu^2\frac{1}{2}H^2+\frac{1}{4}\lambda^2\phi_0^2(H^2+2H^2)+\cdots \\ \nonumber
  =\frac{1}{2}\mu^2H^2+\cdots,
 \end{gather}
we then get from standard derivations of Higgs and gauge boson masses \cite{Cornwell,AitchisonHey4thVol2p380,WeinbergQToFIIpp308,LancasterBlundell2014p437} a Lagrangian for a Higgs field $H/\sqrt{2}$ pertubing around $\phi_0=v/\sqrt{2}$ and a related gauge field $A$
\begin{equation}	\label{eq:LagrangianDimfulFields}
  L=\frac{1}{2}\partial_\mu H\partial^\mu H-\frac{1}{2}\mu^2 H^2+\frac{1}{2}(q\tilde{v})^2A_\mu A^\mu-\frac{1}{4}F_{\mu\nu}F^{\mu\nu}+L_1.
\end{equation}
Here we have hidden interaction terms in $L_1$ and the electric charge coupling constant $q$ originates from the generalized derivative
\begin{equation}
  D_\mu=\partial_\mu+iqA_\mu.
\end{equation}
From the coefficients of the quadratic terms $H^2$ and $A_\mu A^\mu$ in (\ref{eq:LagrangianDimfulFields}) with $q=e=\sqrt{4\pi\alpha}$ and ${\tilde{v}}=\sqrt{2}$ we read off the respective masses $m_H$ and $m_A$ determined by
\begin{equation}	\label{eq:Higgsmass}
  m_Hc^2=\mu\tilde{\Lambda}=\frac{1}{\sqrt{2}}2\pi\frac{\Lambda}{\alpha}=\sqrt{2}(\frac{\pi}{\alpha})^2m_ec^2
\end{equation}
and
\begin{equation}
  m_Ac^2=q\tilde{v}\tilde{\Lambda}=qv=\sqrt{4\pi\alpha}\sqrt{2}\tilde{\Lambda}.
\end{equation}
In (\ref{eq:Higgsmass}) above we have used that the length scale $a$ in the Hamiltonian in (\ref{eq:schroedinger}) relates to the classical electron radius mentioned in the introduction and thus the strong interaction energy scale $\Lambda$ can be conveniently expressed in units of the electron mass $m_e$ by 
\begin{equation} \label{eq:me}
   \Lambda=\frac{\pi}{\alpha}m_ec^2.
\end{equation}
In the neutron decay both an electron and a gauge boson are involved. Thus we use for the gauge mechanism as approximation for $\alpha$ the geometric mean $\tilde{\alpha}$ 
\begin{equation} \label{eq:alphamean}
  {\tilde{\alpha}^{-1}}=1/\sqrt{\alpha(m_e)\alpha(m_Z)}=132.41
\end{equation}
between its known values around electronic energies\cite{RPP2012p107} where $\alpha_0=e^2/(4\pi\epsilon_0\hbar c)=1/137.035999074$ and at bosonic energies\cite{RPP2012p137} where $\alpha_Z=1/127.944$. This yields $m_Hc^2=125.0$ GeV in (\ref{eq:Higgsmass}). See Fig. \ref{fig:HiggsMassGaussians} for a comparison with observations.

The expression (\ref{eq:Higgsmass}) containing solely the electron mass and the fine structure constant and cited again in (\ref{eq:improvednumerics}) is determined by the trailing in (\ref{eq:HiggsAnsatz}) and by the structure of the potential (\ref{eq:periodicpotential}), respectively (\ref{eq:periodicphipotential}) or (\ref{eq:liftedhiggspotential}) and therefore remains valid below. Similarly we would get $m_{A}c^2=78\ \rm{GeV}$. However for the vector gauge field masses corresponding to $m_{A}$ we need to consider the full electroweak $SU(2)_L \times U(1)$ treatment to give the results in (\ref{eq:improvednumerics}).

Note that the usual way of getting the masses for the massive gauge bosons is to derive $v$ from the Fermi coupling constant in muon decay, see e.g. \cite{WeinbergQToFIIp310}, but we use (\ref{eq:HiggsAnsatz}) and (\ref{eq:periodicphipotential}) to give $v=2\pi\sqrt{2}\Lambda/\alpha$ directly in (\ref{eq:tildeenergyscale}) which leads to the values for $m_W$ and $m_Z$ stated in (\ref{eq:improvednumerics}) and following from the standard results in (\ref{eq:mWandmZ}) in the next section.

\section*{7. A full two-component Higgs mechanism}
\label{sec7}

\begin{figure}
\begin{center}
\includegraphics[width=0.45\textwidth]{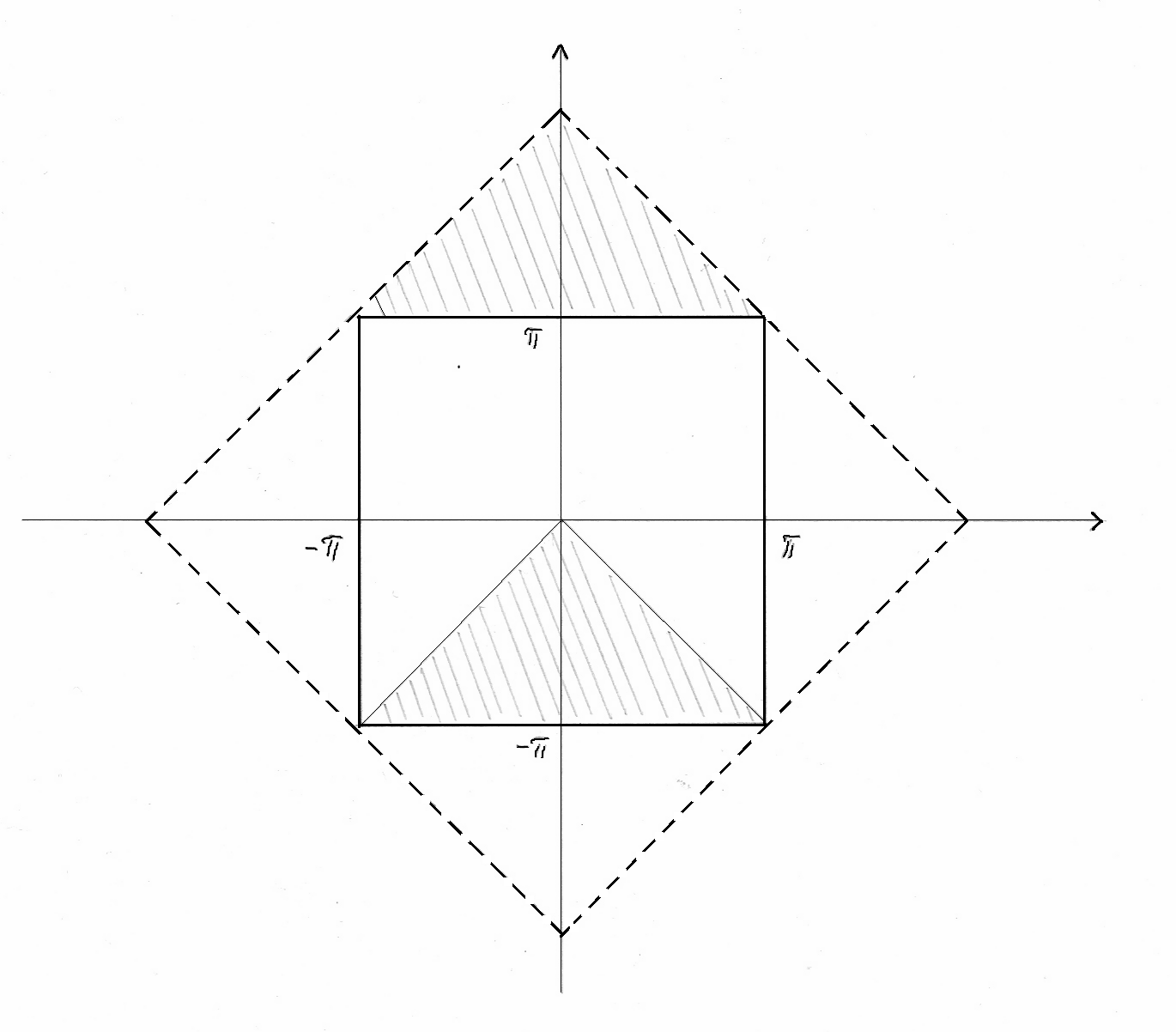}
\caption{A change of variables from the horizontal/vertical $(x,y)$ to a 45 degrees inclined system of coordinates $(u,t)=(\dfrac{x+y}{2},\dfrac{x-y}{2})$ needed in order to find the matrix elements of the centrifugal potential. The seemingly singular denominator in the centrifugal potential (\ref{eq:centrifugalPotential}) is then found to be integrable.  The domain of integration is expanded to suit the new set of variables. This is possible because of the periodicity of the trigonometric functions such that functional values on the hatched area outside the original domain of integration $[-\pi,\pi]\times[-\pi,\pi]$ are identical by parallel transport from the hatched area within that same area.}
\label{fig:variablechange}
\end{center}
\end{figure}

The symmetry breaks introduced by the Bloch phase factors in the parametric eigenstates $g_p$ in (\ref{eq:gp}) have to come in pairs of half odd-integer valued Bloch wave numbers $(\kappa_1, \kappa_2)$ in order to "kill" the singularity in the centrifugal potential
\begin{equation}	\label{eq:centrifugalPotential}
  C=\frac{1}{2}\cdot\frac{4}{3}\sum^3_{ i <  j} \frac{1}{8 \sin^2 \frac{1}{2}(\theta_i -\theta_j)}.
\end{equation}
It namely turns out that the centrifugal potential allows for half-odd-integer \mbox{\boldmath{$\kappa$}}-components provided they come in pairs, for instance \mbox{\boldmath{$\kappa$}} $=(-\frac{1}{2}, \frac{1}{2}, 0)$. In that case we might expand on $g$-couples 
\begin{equation} \label{eq:fpqrpair}
  g_{pqr}-g_{qpr}=e^{ir\theta_3}e^{i(p+q)\frac{\theta_1+\theta_2}{2}}2i\sin\left((p-q)\frac{\theta_1-\theta_2}{2}\right)
\end{equation}
which keep the integrated centrifugal potential regular\cite{Amtrup} because $p-q$ remains integer, see Fig. \ref{fig:variablechange}. 
Here
\begin{equation}
 g_{pqr}=e^{ip\theta_1}e^{iq\theta_2}e^{ir\theta_3}.
\end{equation}
Generalizing the ansatz (\ref{eq:HiggsAnsatz}) we take the paired period doublings corresponding to the shift in Fig. \ref{fig:reducedzone} from $(\kappa_1, \kappa_2)=(0,0)$ to $(\kappa_1, \kappa_2)=(\pm\frac{1}{2},\pm\frac{1}{2})$ to be mediated by a higgs field with a complex two-component doublet $\phi=(\phi_1,\phi_2)$ to "absorb" phase changes (but not kinetic energy nor rest mass) and a two-component electronic lepton $l_{eL}=(\nu_e,e)_L$ to "take care" of  the remaining degrees of freedom (and carry away released energy). Following Cornwell, Aitchison/Hey, Weinberg and Lancaster/Blundell \cite{Cornwell,AitchisonHey4thVol2p380,WeinbergQToFIIpp308,LancasterBlundell2014p437} we then transform to the individual real-valued component vacuum expectation values $<\phi^+>=0$ and $<\phi^0>\equiv\phi_0= v/\sqrt{2}$. In the present case (\ref{eq:tildeenergyscale}) we have
\begin{equation} \label{eq:v}
  v=2\sqrt{2}\frac{\pi}{\alpha}\Lambda
\end{equation}
which relates the electroweak scale to the scale of the strong interactions and which can be inserted into the standard results from the electroweak theory \cite{AitchisonHey4thVol2p381,WeinbergQToFIIp307p309}
\begin{equation} \label{eq:mWandmZ}
  m_Wc^2 =\frac{v|g|}{2}, \ \ m_Zc^2=\frac{v\sqrt{g^2+g'^2}}{2}
\end{equation}
where the $SU(2)$ coupling constant $g$ and the $U(1)$ coupling constant $g'$ are given from the electric charge coupling constant $e=\sqrt{4\pi\alpha}$ and the electroweak mixing angle $\theta_W$ by
\begin{equation} \label{eq:gg'}
  g=-e/\sin\theta_W,\ g'=-e/\cos\theta_W.
\end{equation}
When (\ref{eq:v}), (\ref{eq:me}) and (\ref{eq:gg'}) are used in (\ref{eq:mWandmZ}) together with \cite{RPP2012p107} $\sin^2\hat{\theta}(m_Z)=0.23116$ we can collect our results for the Higgs $m_H$ and the beta decay Fermi coupling constant $G_{F\beta }$ with values for the gauge boson masses $m_W,m_Z$ 
\begin{gather}	
  m_Hc^2=\sqrt{2}\frac{\pi}{\tilde{\alpha}}\Lambda=\sqrt{2}(\frac{\pi}{\tilde{\alpha}})^2m_ec^2=125.0\ \rm{GeV} \nonumber \\  
  \frac{G_{F\beta}}{(\hbar c)^3}=\frac{1}{8\sqrt{2}}(\frac{\tilde{\alpha}}{\pi})^4\frac{1}{(m_ec^2)^2}=1.131\cdot 10^{-5}\ {\rm (GeV)^{-2}} \nonumber \\ 
  m_Wc^2=\sqrt{\frac{4\pi\tilde{\alpha}}{\sin^2\hat{\theta}}}\sqrt{2}(\frac{\pi}{\tilde{\alpha}})^2m_ec^2=80.1\ \rm{GeV} \nonumber \\
  m_Zc^2=\sqrt{\frac{4\pi\tilde{\alpha}}{\sin^2\hat{\theta}\cos^2\hat{\theta}}}\sqrt{2}(\frac{\pi}{\tilde{\alpha}})^2m_ec^2= 91.4\ \rm{GeV}.
   \label{eq:improvednumerics}
\end{gather}
We should stress, that the derivations leading to the Higgs mass in (\ref{eq:Higgsmass}) and (\ref{eq:improvednumerics}) were posted on the preprint archive \cite{TrinhammerHiggsPreprint} prior to newer announcements from The CMS Collaboration \cite{CMSjul2014} and The ATLAS Collaboration \cite{ATLASjun2014}.
The Higgs mass should be compared to these experimental values\cite{CMSjul2014,ATLASjun2014} around $m_Hc^2=125 \ \rm GeV$, namely $124.70\pm 0.34$ GeV and $125.36\pm 0.41$ GeV respectively. The weighted average \cite{TaylorWeightedAverage} of the CMS and ATLAS results is $124.97\pm 0.26\ \rm GeV$. In Fig. \ref{fig:HiggsMassGaussians} for the theoretical result we have used as standard deviation the difference of $0.062\ \rm GeV$ between the Higgs mass from the expression in (\ref{eq:improvednumerics}) with either $\alpha(m_Z)$ or the sliding scale estimate\cite{LandauLifshitzVol4p603,RPP2012p136,WeinbergQToFIIp158and126} for $\alpha(m_H)$ in (\ref{eq:alphamean}). The latter gives $m_Hc^2=124.986\ \rm GeV$ in stead of $m_Hc^2=125.048\ \rm GeV$ from (\ref{eq:alphamean}).

The beta decay Fermi coupling constant in (\ref{eq:improvednumerics}) is related to the muon decay Fermi coupling constant\cite{RPP2012p107} $G_{F\mu}=1.1663787\cdot 10^{-5}\ {\rm (GeV)^{-2}}$  by the quark flavor mixing matrix element\cite{RPP2012p852} $V_{ud}=0.97425$. Thus we compare the result in (\ref{eq:improvednumerics}) with \cite{AitchisonHey4thVol2p235} $G_{F\beta}=G_{F\mu} V_{ud}=1.136\cdot 10^{-5}\ {\rm (GeV)^{-2}}$. The two latter results in (\ref{eq:improvednumerics}) are to be compared with the experimental values\cite{RPP2012p107} $m_Wc^2=80.385(15)\ \rm{GeV}$ and $m_Zc^2=91.1876(21)\ \rm{GeV}$.

Note that the Fermi coupling constant for beta decay has become a derived quantity
\begin{equation} \label{eq:GF}
  \frac{G_{F\beta}}{(\hbar c)^3}=\frac{1}{\sqrt{2}}\frac{1}{v^2}=\frac{1}{8\sqrt{2}}(\frac{\alpha}{\pi})^4\frac{1}{(m_ec^2)^2},
\end{equation}
and that our value for $v$ differs from the standard model edition by a factor\cite{RPP2012p852} $\sqrt{V_{ud}}=\sqrt{0.97425(22)}$. Thus from (\ref{eq:v}) and (\ref{eq:me}) we would find the standard model value $v_{\rm SM}$ of the electroweak energy scale as
\begin{equation}
  v_{\rm SM}=v\sqrt{V_{ud}}=2\sqrt{2}(\frac{\pi}{\tilde\alpha})^2m_ec^2\sqrt{V_{ud}}=246.85\ \rm GeV
\end{equation}
for a geometric mean fine structure constant (\ref{eq:alphamean}). The numerical result is close to the established value\cite{RPP2012p136} $v_{SM}=246.22\ \rm GeV$. Note that (\ref{eq:Higgsmass}), (\ref{eq:v}) and (\ref{eq:improvednumerics}) would give the same result for the Wilson-inspired potential with its $w_{\rm Wilson}(\theta)=1-\cos\theta$ because $w_{\rm Manton}=w$ in (\ref{eq:periodicpotential}) and $w_{\rm Wilson}$ share the mass term $\frac{1}{2}\theta^2$ and the $2\pi$-periodicity. Baryonic states, however, will be shifted downwards by some 20 percent for states constructed from the lowest levels. This would spoil the agreements in Fig. \ref{fig:NDeltaSpectrum}.

\section*{7.1. On the influence of loop corrections on the Higgs mass}

The electroweak scale (\ref{eq:v}) follows from the $2\pi$-shift to a neighbouring trough in (\ref{eq:periodicphipotential}) scaled by the balancing trailing ansatz (\ref{eq:HiggsAnsatz}) in which the coupling constant $\alpha$ appears twice, namely explicitly as a factor on the Higgs field $\phi$ and hidden in the factor $\Lambda$ on the color angle $\theta$. The scale $v$ is settled by the $2\pi$-shift giving a dimensionless $\tilde{v}=\sqrt{2}$ and the Higgs mass is settled by a dimensionless value $\rm{m_H}=1/\sqrt{2}$ from matching the shape of the Higgs potential to that of the neighbouring trough (\ref{eq:periodicphipotential}) of the intrinsic geodetic potential in (\ref{eq:periodicpotential}). With this non-pertubative procedure we have condensed higher loop corrections into the question of finding the right value for the effective coupling constant $\alpha_{\rm eff}$. Pertubative corrections are then contained in the values of $\alpha_{\rm eff}$ in the various steps where they enter. 

The neutron decay, which we have taken to shape the higgs potential, involves both lepton dynamics and gauge boson dynamics wherefore we used in (\ref{eq:alphamean}) for the gauge mechanism as approximation for $\alpha$ the geometric mean
\begin{equation} \label{eq:alphameanNonReciprocal}
  {\tilde{\alpha}}=\sqrt{\alpha(m_e)\alpha(m_Z)}=1/132.41
\end{equation}
between its known values around electronic energies where \cite{RPP2012p107} $\alpha_0=e^2/(4\pi\epsilon_0\hbar c)=1/137.035999074$ and at bosonic energies where \cite{RPP2012p137} $\alpha_Z=1/127.944$. With this we obtained for the Higgs $m_H$, the beta decay Fermi coupling constant $G_{F\beta }$ and the gauge boson masses $m_W,m_Z$ the values in (\ref{eq:improvednumerics}). 

To investigate the influence of fermionic loop corrections on the result for the Higgs, we use expressions from an older work by Jegerlehner \cite{Jegerlehner}. The definition of the total correction $\Delta\alpha$ at scale $\sqrt{s}$ is given by
\begin{equation}
  \alpha(s)=\frac{\alpha}{1-\Delta\alpha(s)}.
\end{equation}

\begin{widetext}

From Jegerlehner we then refer the lepton contributions
\begin{equation}	\label{eq:deltaalphaLepton}
  \Delta\alpha_{\rm leptons}(s)=\sum_{l=e,\mu,\tau}\frac{\alpha}{3\pi}\left[-\frac{8}{3}+\beta_l^2-\frac{1}{2}\beta_l(3-\beta_l^2)\ln(\frac{1-\beta_l}{1+\beta_l})\right],
\end{equation}
where $\beta_l=\sqrt{1-4m_l^2/s}$. The hadronic contributions are
\begin{equation}	\label{eq:deltaalphaHadron}
  \Delta\alpha_{\rm had}^{(6)}(s)=-\frac{\alpha}{9\pi}(1+\alpha_s/\pi)[h(y_d)+h(y_s)+h(y_b)+4(h(y_u)+h(y_c)+h(y_t))],
\end{equation}
where $\alpha_s$ is an effective strong coupling constant and
\begin{equation}
  h(y)=5/3+y-(1+y/2)g(y)
\end{equation}
with $y_i=4m_i^2/s$ and
\begin{equation}
  g(y)=2\sqrt{y-1}\arctan(1/\sqrt{y-1}) \ \ {\rm for} \ \ y>1 
\end{equation}
whereas
\begin{equation}
  g(y)=\sqrt{1-y}\ln(|\frac{1+\sqrt{1-y}}{1-\sqrt{1-y}}|) \ \ {\rm for} \ \ y<1.
\end{equation}

For consistency we use Jegerlehner's set of effective parameters fitted for $E>40\ \rm GeV$ in the above $O(\alpha_s)$ pertubative QCD formula. Jegerlehner states the effective quark masses, $m_{u,d,s,c,b}=0.067,0.089,0.231,1.299,4.500\ \rm GeV$ and the effective $\alpha_s=0.102$. To this we add the top pole value \cite{RPP2014mtop} $m_t=176.7\ \rm GeV$ as a generalization of Jegerlehner's $\Delta\alpha_{\rm had}^{(5)}$. With the lepton masses in (\ref{eq:deltaalphaLepton}) and the effective parameters in (\ref{eq:deltaalphaHadron}) we get, for the scale $\sqrt{s}=125\ \rm GeV$ at the Higgs mass, the following corrections $\Delta\alpha_{\rm leptons}(m_H)=0.032036...$ and $\Delta\alpha_{\rm had}^{(6)}(m_H)=0.029986...$. These sum up to $\Delta\alpha_{\rm fermions}(m_H)=0.0620$ to be compared with the value $\Delta\alpha_{\rm fermions}(m_Z)=0.0592$ from a similar calculation at $\sqrt{s}=91.1876\ \rm GeV$. 

For the radiative corrections we integrate the renormalization group equation for the electric charge coupling constant $e_\mu$
\begin{equation}
 \mu\frac{\partial}{\partial\mu}e_{\mu}=\beta(e_{\mu})
\end{equation}
with the beta-function \cite{WeinbergQToFIIbetafunc}
\begin{equation}
 \beta(e)=\frac{e^3}{12\pi^2}+\frac{e^5}{64\pi^2}+O(e^7),  
\end{equation}
where $\mu$ is the sliding scale. Omitting $O(e^7)$ we get with $b_1=1/(12\pi^2)$ and $b_2=1/(64\pi^2)$
\begin{equation}	\label{eq:renormalResult}
 \ln\mu+k=-\frac{b_2}{b_1^2}\ln e-\frac{1}{2}\frac{1}{b_1}\frac{1}{e^2}+\frac{1}{2}\frac{b_2}{b_1^2}\ln(b_1+b_2e^2)\equiv F(e_\mu),
\end{equation}
where $k$ is an integration constant which in Weinberg's lower order result leads to an expresssion
\begin{equation}
  \alpha_\mu^{-1}=\alpha^{-1}[1-\frac{\alpha}{3\pi}(\ln(\frac{\mu^2}{m_e^2})-\frac{5}{3})].
\end{equation}
This would correspond to a radiative correction
\begin{equation}
  \Delta\alpha_{rad}=\frac{\alpha_0}{3\pi}(\ln(\frac{\mu^2}{m_e^2})-\frac{5}{3})
\end{equation}
with the fine structure constant $\alpha_0=e^2/4\pi=1/137.035999074$. 

We want, however, to go to higher order, but are on the other hand only interested in relations between different scales. So from (\ref{eq:renormalResult}) we continue with the following identity
\begin{equation}
 F(e_\mu)-\ln\mu=F(e_\Lambda)-\ln\Lambda
\end{equation}
relating radiative corrections to $\alpha$ at two different scales $\mu$ and $\Lambda$ to yield
\begin{equation}	\label{eq:alphaslidingHigherOrder}
 \alpha_\mu^{-1}=\alpha_\Lambda^{-1}-\frac{1}{3\pi}\ln(\frac{\mu^2}{\Lambda^2})/(1-\frac{3\pi/4}{\alpha_\Lambda^{-1}+3\pi/4}), \ \ \alpha_\mu,\alpha_\Lambda<<1\ \ {\rm and}\ \ |\frac{\alpha_\Lambda}{\alpha_\mu}-1|<<1,
\end{equation}
i. e. for scales that are not too far apart.

Using the above fermionic corrections together with the sliding scale result (\ref{eq:alphaslidingHigherOrder}) in the geometric mean $\tilde{\alpha}$ in (\ref{eq:alphamean}) would shift the estimated Higgs mass with 3.5 per mille to $124.61\ \rm GeV$. However, if one trusts the heuristic argument leading to the introduction of the geometric mean $\tilde{\alpha}$, one should stick to the prediction of the Higgs mass from the geometric mean between the lepton and the gauge boson sector. This, because the Higgs particle does not occur as such in the neutron decay. It's mass is "only" derived from the gauge mechanism underlying the decay. One could question, however, whether to use $\alpha(m_W)$ in the geometric mean in stead of $\alpha(m_Z)$ since in the standard description it is $W$ that is involved (virtually) in the neutron decay when a d-quark is transformed into a u-quark. Undertaking similar calculations as above we get $\Delta\alpha_{\rm fermions}(m_W)=0.0581...$ and $\Delta\alpha_{\rm fermions}(m_Z)=0.0592...$ which together with the sliding correction
\begin{equation}
 \frac{1}{3\pi}\ln(\frac{m_W^2}{m_Z^2})/(1-\frac{3\pi}{4\alpha_Z^{-1}+3\pi})=0.02725...
\end{equation}
from (\ref{eq:alphaslidingHigherOrder}) gives $\alpha_{\overline{\rm MS}}^{-1}=128.121...$ to yield $m_Hc^2=125.224\ \rm GeV$. Note that we here take the shift $\Delta\alpha_{\overline{\rm MS}}-\Delta\alpha$ between the modified minimal subtraction scheme and the on-shell renormalization scheme \cite{RPP2014aZ} to be the same at $m_W$ as it is at $m_Z$ (namely $0.007165$). This is of course not completely accurate. Glancing at the expression in ref. \cite{RPP2014aZ} for the shift
\begin{equation}
 \Delta\alpha_{\overline{\rm MS}}(m_Z)-\Delta\alpha(m_Z)=\frac{\alpha}{\pi}\left[\frac{100}{27}-\frac{1}{6}-\frac{7}{4}\ln\frac{m_Z^2}{m_W^2}+O(\alpha_s,\alpha_s^2)\right],
\end{equation}
one would expect $\Delta\alpha_{\overline{\rm MS}}(m_W)-\Delta\alpha(m_W)$ to be slightly larger by the amount
\begin{equation}
 \frac{\alpha}{\pi}\cdot\frac{7}{4}\ln\frac{m_Z^2}{m_W^2}=0.001025
\end{equation}
which obviously disappears when $m_Z$ is replaced by $m_W$. For even higher order corrections one would need to know also the shift of the effective strong coupling $\alpha_s$ from $\alpha_s(m_Z)$ to $\alpha_s(m_W)$. We postpone this for future study and refer the interested reader to ref. \cite{Baikov2012aHighLoop}. Here we simply note that already our somewhat crude estimate $\alpha_{\overline{\rm MS}}^{-1}(m_W)=128.12...$ agrees with the Particle Data Group remark \cite{RPP2014aW}, that $\alpha\approx 1/128$ at $Q^2\approx m_W^2$.

\end{widetext}

Cautiously we state our result as a prediction from $m_Z$ with the estimates from $m_W$ and $m_H$ cited as systematic errors. With the newest value \cite{RPP2014aZ} of $\alpha_{\overline{\rm MS}}^{-1}(m_Z)=127.940\pm 0.014$ used in $\tilde{\alpha}=\sqrt{\alpha(m_e)\alpha(m_Z)}$ we arrive at the resulting value $m_Hc^2=125.048\pm0.014({\rm stat.})_{-0.44}^{+0.18}({\rm syst.})\ \rm GeV$.

\begin{figure}
\begin{center}
\includegraphics[width=0.45\textwidth]{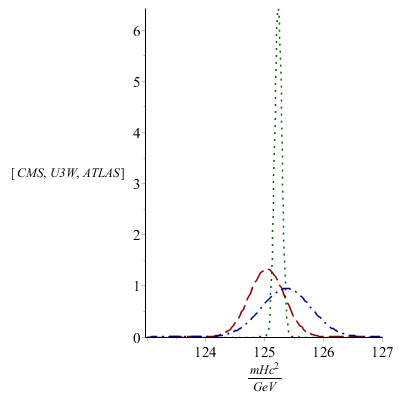}
\caption{Gaussian Higgs mass distributions as observed by the CMS collaboration \cite{CMSfinalRun1} (dashed) and the ATLAS collaboration \cite{ATLASjun2014} (dashdotted) compared with the theoretical result (dotted) in (\ref{eq:improvednumerics}) based on a geometric mean coupling constant between lepton and $W$ gauge boson dynamics. The curve widths represent the standard deviations of the respective mass peak determinations and not the resonance width which is much smaller \cite{CaolaMelnikov2013}. For ease of comparison we have shown the theoretical result with the same standard deviation as in Fig. \ref{fig:HiggsMassGaussians}}
\label{fig:GaussianHiggsW}
\end{center}
\end{figure}

Added in proof we note that the final result from the CMS collaboration \cite{CMSfinalRun1} on Run 1 at the LHC states $m_Hc^2=125.03_{-0.31}^{+0.29}\ \rm GeV$. For completeness we compare in Fig. \ref{fig:GaussianHiggsW} a prediction based on $\alpha_{\overline{\rm MS}}^{-1}(m_W)=128.12...$ with this final CMS-result together with the final ATLAS-result both from the Large Hadron Collider at CERN in Gevena.

\section*{8. Remarks on physical interpretations}
\label{sec8}

The conceptual framework is not the standard model although many aspects comply with it. A benefit is the reduction in the number of ad hoc parameters while keeping - and in certain cases improving on - the agreements with experimental observations. This suggests the framework to be taken as more than just an approximation.

1. The physical conception of baryon dynamics is that of {\it introtangled} energy-momentum with baryons described as stationary states on an intrinsic, compact configuration space. We consider the intrinsic dynamics to be fully described by a Hamiltonian (\ref{eq:schroedinger}) on the intrinsic configuration space, i.e. not as fields of quarks and gluons in laboratory space with separate strong and electroweak interaction parts. Rather we consider the baryons to be entire entities of introtangled energy-momentum which carry strong and electroweak manifestations intermingled. Quarks (\ref{eq:quarkdR}) and gluons (\ref{eq:gluondPhi}) come about when the intrinsic states are projected to laboratory space. In the language of the standard model we have confinement per construction since we take the configuration space to be compact.

2. We consider the creation of electric charge to originate in topological changes (\ref{eq:gp}) in the intrinsic states, see also Fig. \ref{fig:reducedzone}. As configuration space we take the Lie group $U(3)$. It contains as intermingled subspaces exemplars of both $U(1)$, $SU(2)$ and $SU(3)$ structures, e.g. the gauge group $SU(3)$ of strong interactions and the gauge group $U(1)\times SU(2)$ of the electroweak interactions.

3. We consider the strong and electroweak energy scales to be related by a balancing of color and higgs field energies (\ref{eq:HiggsAnsatz}) in the weak decay of baryons. We take the length scale of the strong interaction sector of the model to be settled in the projection of the neutron decay which relates changes in the intrinsic baryon states to the electroweak sector (\ref{eq:periodicphipotential}), (\ref{eq:tildeenergyscale}), (\ref{eq:v}). We thus take a projection of the intrinsic geometry to the electrically defined, classical electron radius as an input for the strong interaction scale with the electron imagined as a "peel off" from the neutron, leaving a "charge scarred" proton torus, see Fig. \ref{fig:ProjectionGroupAlgabra}. Further we use a trailing ansatz (\ref{eq:HiggsAnsatz}) to relate strong and electroweak coupling constants in order to set the scale (\ref{eq:v}) for the electroweak sector and its Higgs (\ref{eq:Higgsmass}) and gauge boson masses (\ref{eq:mWandmZ}).

4. States are projected from intrinsic space to laboratory space by use of the exterior derivative, the momentum form on the intrinsic manifold (\ref{eq:exteriorderivativeDefinition}), respectively (\ref{eq:quarkdR}) and (\ref{eq:gluondPhi}). From projection of the intrinsic structure to space we recognize the toroidal generators as momentum operators (\ref{eq:toroidalParametricMomentum}) and off-torus generators as spin and flavor operators (\ref{eq:KandMcommutation}), (\ref{eq:M2spectrum}). In experimental production of resonances we see from space: {\it The impact momentum generates the abelian maximal torus of the $U(3)$ intrisic space. The momentum operators act as introtangling generators}. When decay, asymptotic freedom, fragmentation and confinement are of concern we see from intrinsic space: {\it The quark and gluon fields are projections of the vector fields induced by the momentum form on the intrinsic states}. The projected fields are treated as quantum fields and a balancing trailing ansatz between color and Higgs field energies in weak baryon decays connects strong and electroweak sectors (\ref{eq:HiggsAnsatz}) via the period doublings (\ref{eq:fpqrpair}) allowed in the parametrization of the intrinsic space. The structure of the period doublings and the intrinsic potential (\ref{eq:periodicpotential}) determines the Higgs potential from which the Higgs mass originates.

5. Because the dynamical structure is formulated on the Lie group, it will show different manifestations depending on which derivatives (\ref{eq:exteriorderivativeDefinition}) one is taking. For instance we interpret the three toroidal dimensions as intrinsic color quark degrees of freedom (\ref{eq:dPhi}), (\ref{eq:psiVector}). These are intermingled with flavor degrees of freedom. And both are intermingled with the eight gluon dimensions laid out by the Gell-Mann matrices (\ref{eq:gluondPhi}). Thus we do not consider color and flavor degrees of freedom as being independent. As mentioned in Subsec. 2.2, the distribution functions in Fig. \ref{fig:partonDistribution} are produced by using the exterior derivative (\ref{eq:dPhi}) on tracks \cite{TrinhammerEPL102} from the quark flavor generators $T_u=2/3\ T_1-T_3$ and $T_d=-1/3\ T_1-T_3$. And the reduction in the number of independent quark degrees of freedom practically eliminates the missing resonance problem in ordinary QMs, compare Figs. \ref{fig:NDeltaSpectrum} and \ref{fig:missingResonances}.

\section*{9. Examples for future study}
\label{sec9}

The neutral flavor, neutral electric charge baryon singlets mentioned in Sec. 4 and listed in Table \ref{tab:singlets} should be sought for. They may even lie dormant in data pools already taken since they have no charged partners to help them surface in partial wave analysis.

A more accurate estimate of the coupling constant $\alpha_{\overline{\rm MS}}(m_W)$ at $W$-bosonic energies is wanted for an even more accurate prediction of the Higgs mass $m_H$.

A suitable base on which to expand for exact solutions for charged baryons is wanted in order to improve the predictions on the N and $\Delta$ mass spectrum.

The geodetic distance potential (\ref{eq:mantonTracePotential}) can be used as an interaction term. For instance in a model for two baryons with configuration variables $u$ and $u'$ for which $d(u,u')=d(e,u^\dagger u')$. Thus we conjecture the deuteron to be the spin 1 ground state of
\begin{equation}
   \frac{\hbar c}{a}\left[-\frac{1}{2}\Delta_u-\frac{1}{2}\Delta_{u'}+\frac{1}{2}d^2(u,u')\right]\Psi(u,u')=E\Psi(u,u').
\end{equation}
When one imagines a projection of the term $u^\dagger u'$ it has an antiquark-quark structure characteristic of mesons in that the $u^\dagger$ when projected is to be represented on an antiquark to the left and the $u'$ is to be represented on a quark to the right.

\section*{10. Conclusion}
\label{sec10}

We have derived the Higgs mass and the electroweak energy scale by connecting structurally the strong and electroweak baryon sector. We have considered baryons as entire entities on an intrinsic $U(3)$ configuration space with a hamiltonian structure to yield baryon mass spectra. The parametrization of the intrinsic baryon space and its potential allows for period doublings which determines the Higgs potential and settles the Higgs mass. Parton distribution functions follow from the exterior derivative, the momentum form on intrinsic states.

The general agreement of the various derivations with experimental observations suggests further investigations within the model. In particular a base for exact solutions of electrically charged baryonic states is wanted as well as experimental investigations looking for neutral flavor, neutral charge baryon singlets particular for the present model. The  singlets should be visible as resonances in negative pions scattering on protons, in photoproduction on neutrons, in neutron diffraction dissociation experiments and in invariant mass spectra of protons and negative pions in B-decays. The Higgs mass prediction, the singlet predictions and the elimination of a missing baryon resonance problem distinguish the present model from the standard model predictions. We await singlet searches on GeV-machines from new experiments or from dedicated analysis on existing data pools and we await more accurate Higgs mass measurements from Run 2 at the Large Hadron Collider.

\section*{Acknowledgments}
One of us (O. L. Trinhammer) would like to express his thanks to a long list of people: Anonymous referees for clarifying questions. My teachers Geoffrey C. Oades and Abel Miranda for inspiration.  I thank Jeppe Dyre, Peter H. Hansen, Anders Andersen, Jakob Bohr, Tomas Bohr, J\o rgen Kalckar and Victor F. Weisskopf for advice. I thank for helpful discussions and lending ear: Torben Amtrup, Holger Bech Nielsen, J\o rgen Beck Hansen, J\o rn Dines Hansen, Vladimir B. Kopeliovich, Svend Bj\o rnholm, Bo-Sture Skagerstam, Per Salomonson, Mikul\'a\v{s} Bla\v{z}ek, P. Filip, \v{S}tefan Olejn\'ik, M. Nagy, A. Nogov\'a, Peter Pre\v{s}najder, Vladim\'ir \v{C}ern\'y, Juraj Boh\'a\v{c}ik, Roman Lietava, Vracheslav P. Spiridonov, Ben Mottelson, Andreas Wirzba, Niels Kj\ae r Nielsen, Pavol Valko. I thank for help with the Laplacian and numerical methods: Gestur Olafsson, Yurii Makeenko, Dmitri Boulatov, Karen Ter-Martirosyan, Hans Bruun Nielsen, Jens Hugger, Kurt Munk Andersen, Hans Plesner Jacobsen, Povl Holm, Karsten Wedel Jacobsen. I thank for technical help: Erik Both, Bjarne Bach, Knud Fjeldsted and Jens Bak. I thank for encouragement and inspiration: Mads Hammerich, Poul Werner Nielsen, Norbert Kaiser, Jaime Vilate, Pedro Bucido, Manfried Faber, Karl Moesgen, A. Di Giacomo, Ivan \v{S}tich, Benny Lautrup, Poul Olesen,  Miroslava Smr\v{c}inov\'a, Lissi Regin, Berit Bj\o rnow, Elsebeth Obbekj\ae r Petersen, Hans Madsb\o ll. I thank for institutional framework: Leo Bresson, Mogens Hansen, Jane Hvolb\ae k Nielsen.


\section*{{References}}


\begin{thebibliography}{99}

\bibitem{EnglertBrout}
 F. Englert and R. Brout, Phys. Rev. Lett. {\bf 13}(9) (1964) 321-323.
\bibitem{HiggsSep1964}
 P. W. Higgs, Phys. Lett. {\bf 12}(2) (1964) 132-133.
\bibitem{HiggsOct1964}
 P. W. Higgs, Phys. Rev. Lett. {\bf 13}(16) (1964) 508-509.
\bibitem{GuralnikHagenKibble}
 G. S. Guralnik, C. R. Hagen, T. W. B. Kibble, Phys. Rev. Lett. {\bf 13}(20) (1964) 585-587.
\bibitem{Higgs1966}
 P. W. Higgs, Phys. Rev. {\bf 145}(4) (1966) 1156-1163.
\bibitem{ATLASsep2012}
 ATLAS Collaboration, Phys. Lett. B {\bf 716} (2012) 1-29. ArXiv:1207.7214v1 [hep-ex].
\bibitem{CMSsep2012}
 CMS Collaboration, Phys. Lett. B {\bf 716} (2012) 30-61. ArXiv:1207.7235v1 [hep-ex].
\bibitem{CMSjul2014}
  CMS Collaboration, arXiv:1407.0558v1 [hep-ex].
\bibitem{ATLASjun2014}
  ATLAS Collaboration, Phys. Rev. D {\bf 90} (2014) 052004, arXiv:1406.3827v1 [hep-ex].
\bibitem{RPP2012CKMmatrix}
 Particle Data Group (J. Beringer et al.), Phys. Rev. D {\bf 86} (2012) 010001, p. 157.
\bibitem{FlorianScheckU2}
  F. Scheck, {\it Electroweak and Strong Interactions. Phenomenology, Concepts, Models}, $3^{\rm rd}$ ed., (Springer-Verlag, Berlin, Heidelberg, 1996, 2012), p. 232.
\bibitem{Schwartz2014}
  M. D. Schwartz, {\it Quantum Field Theory and the Standard Model}, (Cambridge University Press, Cambridge, UK 2014), p. 161.
\bibitem{CaolaMelnikov2013}
  F. Caola and K. Melnikov, Phys. Rev. D {\bf 88} (2013) 054024, arXiv: 1307.4935v3 [hep-ph].
\bibitem{DharMandalWadia}
 A. Dhar, G. Mandal and S. R. Wadia, Phys. Rev. D {\bf 80} (2010) 105018. ArXiv: 0905.2928v3 [hep-th] (2009).
\bibitem{GrossNeveu}
 D. J. Gross and A. Neveu, Phys. Rev. D {\bf 10}, 3235 (1974).
\bibitem{NambuJonaLasinio}
 Y. Nambu and G. Jona-Lasinio, Phys. Rev. {\bf 122}, 345 (1961).
\bibitem{TrinhammerEPL102}
 O. L. Trinhammer, Eur. Phys. Lett. {\bf 102} (2013) 42002. ArXiv:1303.5283v2 [physics.gen-ph].
\bibitem{TrinhammerHiggsPreprint}
 O. L. Trinhammer, arXiv: 1302.1779v2 [hep-ph].
\bibitem{KogutSusskind}
 J. B. Kogut and L. Susskind, Phys. Rev. D {\bf 11}(2) (1975) 395.
\bibitem{Manton}
 N. S. Manton, Phys. Lett. B {\bf 96} (1980) 328-330.
\bibitem{Trinhammer1983}
 O. Trinhammer, Phys. Lett. B {\bf 129}(3,4) (1983) 234-238.
\bibitem{RPP2012p131}
 Particle Data Group (J. Beringer et al.), Phys. Rev. D {\bf 86} (2012) 010001, p. 131.
\bibitem{WeinbergQToFIIp186}
 S. Weinberg, {\it The Quantum Theory of Fields - Modern Applications}, Vol. II, (Cambridge University Press 1995/2012), p. 186.  
\bibitem{BohrProvidenciaProvidencia2005}
 H. Bohr, C. Providencia and J. da Providencia, Phys. Rev. C {\bf 71} (2005) 055203.
\bibitem{DiakonovPetrovPolyakov}
 D. Diakonov, V. Petrov and M. Polyakov, Z. Phys. A {\bf 359} (1997) 305-312. ArXiv:9703373v2 [hep-ph].
\bibitem{Heisenberg}
 W. Heisenberg, Ann. d. Phys. (5) {\bf 32} (1938) 20-33.
\bibitem{LandauLifshitz}
 L. D. Landau and E. M. Lifshitz, {\it The Classical Theory of Fields, Course of Theoretical Physics} Vol. 2,  4$^{\rm{th}}$ ed., (Elsevier Butterworth-Heinemann, Oxford 2005), p. 97.
\bibitem{RPP2012p107}
 See ref. \cite{RPP2012p131} p. 107.
\bibitem{RPP2012p137}
 See ref. \cite{RPP2012p131} p. 137!
\bibitem{Wilson}
 K. G. Wilson, Phys. Rev. D {\bf 10} (1974) 2445.
\bibitem{Wadia}
 S. R. Wadia, Phys. Lett. B {\bf 93}(4) (1980) 403-410.
\bibitem{HansPlesnerJacobsen}
 H. P. Jacobsen, Department of Mathematics, University of Copenhagen, private communication approx. 1997.
\bibitem{RPP2012}
 See ref. \cite{RPP2012p131} pp. 78-83.
\bibitem{Hoehler}
 G. H\"{o}hler in Particle Data Group, Phys. Lett. {\bf B239}, (1990), p. VIII.10.
\bibitem{TrinhammerOlafsson}
 O. L. Trinhammer and G. Olafsson, arXiv:9901002v2 [math-ph].
\bibitem{Weyl}
 H. Weyl, {\it The Classical Groups - Their Invariants and Representations}, 2$^{\rm{nd}}$ ed. (Princeton University Press 1997), p. 197. 
\bibitem{SchiffGellMannMatrices}
 L. I. Schiff, {\it Quantum Mechanics}, 3rd. ed., (McGraw-Hill, 1955/1968), p. 210.
\bibitem{Okubo}
 S. Okubo, {\it{Note on Unitary  Symmetry in Strong Interactions}}, Prog. Theor. Phys. {\bf{27(5)}}, 949-966, (1962). 
\bibitem{GellMann1962}
 M. Gell-Mann, {\it{Symmetries of Baryons and Mesons}}, Phys.\ Rev.\ {\bf{125(3)}}, 1067-1084, (1962).
\bibitem{Neeman}
 Y. Ne'eman, {\it{Derivation of Strong Interactions from a Gauge Invariance}}, Nucl. Phys.{\bf{26}}, 222-229, (1962).
\bibitem{Gasiorowicz}
 S. Gasiorowicz,{\it{Elementary Particle Physics}} (Wiley and Sons, New York, 1966), p.287.
\bibitem{Bruun}
 H. Bruun Nielsen, Technical University of Denmark, (suggested in private communications, 1997).
\bibitem{Dowker}
 J. S. Dowker, Ann. of Phys. {\bf{62}} (1971) 361-382.
\bibitem{Slater}
 J. C. Slater, Phys. Rev. {\bf 34} (1929) 1293.
\bibitem{LandauLifshitzVol4p603}
 V. B. Berestetskii, E. M. Lifshitz and L. P. Pitaevskii, {\it Quantum Electrodynamics - Landau and Lifshitz Course of Theoretical Physics}, Vol 4, $2^{nd}$ edition, Elsevier (1982/2004), p. 603.
\bibitem{RPP2012p136}
 See ref. \cite{RPP2012p131} p. 136.
\bibitem{WeinbergQToFIIp158and126}
 See ref. \cite{WeinbergQToFIIp186} p. 158 and p. 126.
\bibitem{BaryonLieProgrammes}
 Programmes available at \url{https://dcwww.fysik.dtu.dk/~trinham/BaryonLieProgrammes/}.
\bibitem{AshcroftMermin}
 N. W. Ashcroft and N. D. Mermin, {\it Solid State Physics}, (Holt, Rinehart and Winston, New York 1976), p. 160. 
\bibitem{RPP2012p205}
 See ref. \cite{RPP2012p131} p. 205. 
\bibitem{RPP2012p79}
 See ref. \cite{RPP2012p131} p. 79.
\bibitem{PovlHolm}
 P. Holm, Rungsted Gymnasium, Denmark, private communication 1993.
\bibitem{KlemptRichard}
 E. Klempt and J. M. Richard, Rev.\ Mod.\ Phys.\ {\bf{82(2)}} (2010) 1095.
\bibitem{Aleev}
 A. N. Aleev et al., Z. Phys. C {\bf{25}} (1984) 205-212.
\bibitem{Epecur}
 EPECUR Collaboration and GW INS Data Analysis Center (I. G. Alekseev et al.), arXiv:1410.6418v1 [nucl-ex].
\bibitem{Zhu}
 L. Y. Zhu et al., Phys. Rev. C {\bf{71}} (2005) 044603.
\bibitem{Babar}
 BABAR Collaboration (B. Aubert et al.), Phys. Rev. D {\bf{79}} (2009) 112009.
\bibitem{CornwellAppendixJ}
 J. F. Cornwell, {\it Group Theory in Physics}, Vol. 2, (Elsevier Academic Press, Amsterdam 1984/2004), Appendix J.
 \bibitem{HolgerBechNielsen}
 H. Bech Nielsen, Niels Bohr Institute, Copenhagen, private communication 199x.
\bibitem{GuilleminPollack}
 V. Guillemin and A. Pollack, {\it Differential Topology}, (Prentice-Hall, New Jersey, USA 1974), p. 163.
\bibitem{BrodskyEtAlPhaseFactor}
 S. J. Brodsky, H.-C. Pauli and S. S. Pincky, Phys. Rep. {\bf 301}, 299-486 (1998), p. 312.
\bibitem{RPP2004}
 Particle Data Group (S. Eidelman et al.), Phys. Lett. B {\bf 592} (2004).
\bibitem{DonoghueEtAlGaugeTransformation}
 J. F. Donoghue, E. Golowich and B. R. Holstein, {\it Dynamics of the Standard Model} (Cambridge University Press, Cambridge 1992/1996), p. 16.
\bibitem{GriffithsElementaryParticlesp403} 
  D. Griffiths, {\it Introduction to Elementary Particles}, $2^{nd}$ ed., (Wiley-VCH, Weinheim, Germany 2008/2012) p. 403.
\bibitem{GriffithsElementaryParticlesp355}
 See ref. \cite{GriffithsElementaryParticlesp403} p. 355.
\bibitem{WeinbergQToFIIp303}
 See ref. \cite{WeinbergQToFIIp158and126} p. 303.
\bibitem{GriffithsElementaryParticlesp381} 
 See ref. \cite{GriffithsElementaryParticlesp403} p. 381.
\bibitem{AitchisonHey4thVol2p380}
 I. J. R. Aitchison and A. J. G. Hey, {\it{Gauge Theories in Particle Physics - A Practical Introduction}}, 4$^{\rm{th}}$ ed., Vol. 2, (CRC Press, Boca Raton, London, New York 2013), p. 380.
\bibitem{Cornwell}
 See ref. \cite{CornwellAppendixJ} pp. 766, 786, 799.
\bibitem{WeinbergQToFIIpp308}
 See ref. \cite{WeinbergQToFIIp186} pp. 308.
\bibitem{LancasterBlundell2014p437}
 T. Lancaster and S. J. Blundell, {\it Quantum Field Theory for the Gifted Amateur}, (Oxford University Press, Oxford, United Kingdom, 2014), p. 437.
\bibitem{WeinbergQToFIIp310}
 See ref. \cite{WeinbergQToFIIp186} p. 310.
\bibitem{Amtrup}
 T. Amtrup, {\it Two integral presumptions}, LMFK-bladet no. 4, April 1998. Available at \url{https://dcwww.fysik.dtu.dk/~trinham/BaryonLieProgrammes/}.
\bibitem{AitchisonHey4thVol2p381}
 See ref. \cite{AitchisonHey4thVol2p380} p. 381.
\bibitem{WeinbergQToFIIp307p309}
 See ref. \cite{WeinbergQToFIIp186} p. 307, p. 309.
\bibitem{TaylorWeightedAverage}
 J. R. Taylor, {\it An Introduction to Error Analysis}, 2$^{nd}$ ed., (University Science Books, California, USA, 1982/1997), pp. 175.
\bibitem{RPP2012p852}
 See ref. \cite{RPP2012p131} pp. 157, 852.
\bibitem{AitchisonHey4thVol2p235}
 See ref. \cite{AitchisonHey4thVol2p380} p. 235. 
\bibitem{RPP2012p852}
 See ref. \cite{RPP2012CKMmatrix} pp. 157, 852.
\bibitem{HalzenMartinGfbeta}
 F. Halzen and A. J. Martin, {\it Quarks and Leptons: An Introductory Course in Modern Particle Physics}, (J. Wiley and Sons, New York, Chichester, Brisbane, Toronto, Singapore, 1984), p. 282.
\bibitem{WeinbergQToFFIIp185}
 See ref. \cite{WeinbergQToFIIp186} p. 185.
\bibitem{Jegerlehner}
 F. Jegerlehner, Nucl. Phys. Proc. Suppl. {\bf 51} C (1996), 131-141. ArXiv:hep-ph/960684v1.
\bibitem{RPP2014mtop}
 K. A. Olive et al. (Particle Data Group), Chin. Phys. C, 2014, {\bf 38}(9): 090001. p.33.
\bibitem{WeinbergQToFIIbetafunc}
 See ref. \cite{WeinbergQToFIIp186} p. 126.
\bibitem{RPP2014aZ}
 See ref. \cite{RPP2014mtop} p. 140.
\bibitem{Baikov2012aHighLoop}
 P. A. Baikov et al., JHEP, {\bf 1207}, 017 (2012), ArXiv:1206.1284 [hep-th].
\bibitem{RPP2014aW}
 See ref. \cite{RPP2014mtop} p. 109.
\bibitem{CMSfinalRun1}
 CMS Collaboration, arXiv:1412.8662 [hep-ex].



\end{thebibliography}
\end{document}